\documentclass[%
reprint,
superscriptaddress,
nobibnotes,
 amsmath,amssymb,
prb,
twocolumn,
floatfix,
showkeys 
]{revtex4-2}

\usepackage{graphicx}
\usepackage{dcolumn}
\usepackage{bm}
\usepackage[colorlinks,linkcolor=blue,citecolor=magenta,hyperindex,CJKbookmarks]{hyperref}

\usepackage{tabularx} 

\usepackage[export]{adjustbox} 

\usepackage{xcolor}

\newcommand{\dpo}{$A_2BB'$O$_6$}
\newcommand{\bss}{Ba$_2$ScSbO$_6$}
\newcommand{\bst}{Ba$_2$ScTaO$_6$}

\newcommand{\byn}{Ba$_2$YNbO$_6$}
\newcommand{\byt}{Ba$_2$YTaO$_6$}
\newcommand{\css}{Ca$_2$ScSbO$_6$}
\newcommand{\sss}{Sr$_2$ScSbO$_6$}

\begin{document}

\title[Excitons in double perovskite oxides]{Electronic structure fingerprints of visible range excitons in $d^0$ double perovskite oxides}

\author{Bhagyashree Behera}
\altaffiliation[]{These authors contributed equally.}
\affiliation{Discipline of Natural Sciences, Indian Institute of Information Technology, Design and Manufacturing, Khamaria, Jabalpur-482005, India.}

\author{Debatri Ash}
\altaffiliation[]{These authors contributed equally.}
\affiliation{Department of Physics, Birla Institute of Technology Mesra, Ranchi, India - 835215.}

\author{Urmimala Dey}
\email{urmimala.dey@list.lu}
\affiliation{Centre for Materials Physics, Durham University, South Road, Durham DH1 3LE, United Kingdom}
\affiliation{Luxembourg Institute of Science and Technology (LIST), Avenue des Hauts-Fourneaux 5, L-4362, Esch-sur-Alzette, Luxembourg}

\author{M. K. Roy}
\affiliation{Discipline of Natural Sciences, Indian Institute of Information Technology, Design and Manufacturing, Khamaria, Jabalpur-482005, India.}

\author{Pritha Patra}
\affiliation{Specialty Glass Division, CSIR-Central Glass and Ceramic Research Institute, 196, Raja S. C. Mullick Road, Kolkata 700 032, India.}
\affiliation{Academy of Scientific and Innovative Research (AcSIR), CSIR-Human Resource Development Centre, (CSIR-HRDC) Campus, Postal Staff College Area, Sector 19, Kamla Nehru Nagar, Ghaziabad, Uttar Pradesh-201 002, India.}%

\author{K. Annapurna}
\affiliation{Specialty Glass Division, CSIR-Central Glass and Ceramic Research Institute, 196, Raja S. C. Mullick Road, Kolkata 700 032, India.}

\author{S. K. Rout}
\affiliation{Department of Physics, Birla Institute of Technology Mesra, Ranchi, India - 835215.}

\author{Ajay K Himanshu}
\email{akh@vecc.gov.in}
\affiliation{Variable Energy Cyclotron Center, 1/AF Bidhannager, Saltlake, Kolkata, India - 700064.}
\affiliation{
 Homi Bhabha Nattional Institute, Mumbai, India - 400094.}%

\author{Rajyavardhan Ray}
\email{r.ray@bitmesra.ac.in}
\affiliation{Department of Physics, Birla Institute of Technology Mesra, Ranchi, India - 835215.}


\date{}

\begin{abstract}
The presence of excitons significantly influence the optoelectronic properties and potential applications of materials. 
Using combined theoretical and experimental tools, we investigate the absorption spectra of $d^0$ double  perovskite oxides, 
Ba$_{2}$Y$B'$O$_6$ ($B'$ = Nb, Ta, Sb), Ba$_{2}$Sc$B'$O$_6$ ($B'$ = Ta, Sb) and $A_{2}$ScSbO$_6$ ($A$=Ca, Sr, Ba), 
allowing for a systematic variation of composition. We not only show that low-energy excitons possessing large binding energies 
up to 3 eV are present in the visible range in {\it all} the considered wide-gap insulators, but also that the nature 
and properties of these excitons differs from those in double perovskite halides as well as perovskite oxides. We provide 
insights on the origin of such differences by a comparative analysis of the electronic structure. Our findings elucidate possible 
correlations between the exciton properties and the composition, via the electronic structure, towards a comprehensive understanding 
of correlation effects and rational design principles.
\end{abstract}


\maketitle


\section{\label{sec:intro}Introduction}

Wide-gap insulators are considered promising for a variety of optoelectronic and photocatalytic applications \cite{Eng2003,Vasala2015, Ray2017, 
Yin2019, Nair2022,Ji2023}.
In general, varying degree of sensitivity to crystal structure and composition is found, and viability for specific applications is dictated by 
the electronic and optical properties \cite{Eng2003,Ray2017,Yin2019,Nair2022,Ji2023}. The large-gap $d^0$ double perovskite oxides (DPOs), 
general formula $A_2BB'O_6$, formed by transition metals with unfilled/empty valence $d$-shells at the $B$/$B'$-sites are 
considered promising in microwave-dielectric resonator applications, including interference filters,
reflective coating and in optical fibers, due to their good dielectric properties, and as buffer materials due to their
low reactivity \cite{Eng2003,Ray2017,Yin2019}. Within this materials class, a wide range of optoelectronic properties can be tuned by suitable 
modulation of composition \cite{Vasala2015}. For example, a bandgap increase of up to 2 eV can be realized with increasing octahedral tilting and 
distortions \cite{Eng2003,Ray2017}. 

Further, many body effects, such as the presence of bound electron-hole pair quasiparticles --- the so-called excitons, can lead to sub-gap features 
in optical properties. An accurate knowledge of electronic and optical properties, including the details of exciton modes, is thus important.  
The exciton binding energy is also useful to determine charge recombination, 
an important parameter for devices. In addition, these excitons may find applications in excitonic emission \cite{Peters2022}, 
photocatalysis \cite{Wang2024}, energy devices \cite{Jang2018, Hu2023} and even information processing \cite{Datta2022}. As a result, there is a 
growing interest in the study of excitons in a variety of materials due to potential optoelectronic applications \cite{Biega2023, Omelchenko2017}. 
However, systematic studies on detailed characterization of excitons and possible correlations between the electronic properties and the exciton 
characteristics are limited.

An exciton is an electron-hole pair bound together by Coulomb interaction. Formation of excitons in semiconductors and insulators is 
well-known \cite{Lushchik2000, Koch2006, Chiodo2010, Kang2010, Rodl2013, Lany2015, Baldini2017} and occurs upon irradiation by light. 
Due to their binding energy, their presence is often detected by peaks in the electronically forbidden energy range (bandgap) in the absorption 
spectrum \cite{Nikl2000,Laguta2007,Pudewill1976,Grunberg1995,Elliott1957}, and the photoluminiscence spectra \cite{Ito1997,Peters2022}.
Predictions for real materials require density functional (DF) calculations incorporating many body effects \cite{Quintela2022} via the Bethe-Salpeter 
equation (BSE)~\cite{Reining2016,Blase2020} which, unfortunately, is computationally quite demanding. Therefore, reliable predictions for real 
materials are scarce. 

As the metal-ligand bonds and crystal structure largely govern the electronic structure of $d^0$-DPOs \cite{Eng2003,Ray2017}, exciton properties in related systems are noteworthy. 
In certain transition metal oxides with partially occupied valence $d$-states, 
the excitons are spatially localized within a radius of the order of metal-oxygen bonds \cite{Biswas2018}. 
In non-magnetic perovskite oxides with $d^{0}$ configuration, the contributions from the band edge states with dominant $d$-character, depend on the composition and can lead to large 
exciton binding energies up to 6 eV \cite{Varrassi2021}. For example, in SrTiO$_3$ and KTaO$_3$, the excitons are spatially delocalized with 
low-energy transitions to metal $t_{2g}$ states, whereas in SrHfO$_3$ and SrZrO$_3$, transitions to $t_{2g}$ states also contribute at low energies 
while the excitons are spatially localized. In both cases, the dominant contribution corresponds to transitions at the $\Gamma$-point. Deviation 
from cubic structures leads to sizable contributions from different $k$-points as well, leading to spatially localized excitons. 
 
On the other hand, in the double perovskite halides, based on solution to BSE, the nature and the properties of the exciton modes is predicted to be composition 
dependent. In general, both hydrogenic (Wannier-Mott) and non-hydrogenic excitons are present, and the binding energies can be as large as 2 eV \cite{Biega2023}. 
The presence of indirect bandgaps and large effective electron/hole mass anisotropies stabilize non-hydrogenic excitons which are highly delocalized in the reciprocal space.

For $d^0$-DPOs, therefore, a number of questions arise: How do the findings in the perovskite transition metal oxides extend to $d^0$ double perovskites when two different transition 
metals are present? Especially, what happens if both the transition metals contribute at the conduction band (CB) edge? What if the conduction band 
edge does {\it not} have dominant contributions from the metal $d$ states but an extended $s/p$-state instead? What is the role of ligand anions?  
Most importantly, is it possible to extract reliable information about the excitons from the elementary electronic structure based on density functional theory (DFT) alone?

To address these concerns, here, we focus on $d^0$-DPOs --- a versatile playground to explore the effects of bandgap, crystal structure and 
composition \cite{Vasala2015} --- and study the optical response towards establishing 
presence and nature of excitons in the $d^0$-DPOs as a function of composition and structure. We further investigate and identify the key characteristics of 
the electronic strucutre correlated with the observed exciton features. Typically, divalent alkaline earth cations occupy the $A$-sites, while the $B$ and $B'$ sites are occupied by 
transition metal cations with effectively filled or empty $d$-states, leading to an insulating state in {\dpo} \cite{Eng2003,Vasala2015,Ray2017}.
In particular, we consider the Sc and Y based DPOs, $A_2$ScSbO6 ($A$=Ba, Sr, Ca), Ba$_2$Y$M$O$_6$ ($M$ = Ta, Nb, Sb) and Ba$_2$Sc$M$O$_6$ 
($M$ = Ta, Sb), and carry out a systematic and comparative analysis of the absorption spectra aided with DF calculations using the generalized gradient 
approximation (GGA)~\cite{Perdew1996} and modified Becke-Johnson (mBJ) potentials \cite{Tran2009,Koeller2012,Jishi2014}. The observed absorption 
spectra are analyzed using the Kubelka-Munk (KM) function (without excitons), Elliot formula for isolated exciton peaks and excitons at the band 
edge, and the underlying electronic structure. 

We observe multiple low-energy peaks in the absorption spectra identified as exciton peaks. The peaks are of considerably weak intensity and extend 
all the way to the visible range, much below the corresponding bandgaps. Based on a high-quality fit of the low-energy part of the data, we obtain the 
binding energies and linewidths (FWHM) from these peaks. Effects of the crystal structure and composition, in terms of the electronic structure, on the 
origin and nature of these excitons is reported. Systematic variations in the peak shapes is found as the atoms/orbital contributions at the band edges vary.
These findings elucidate the role of $A$-site and $B/B'$-site cations on the sub-gap excitons in the low-energy optical spectra of the considered DPOs.

\section{Results and Discussions}

\subsection{Crystal Structures \& Electronic Properties}

Structurally ordered DPOs, $A_2BB'$O$_6$, consist of transition metals $B$ and $B'$ surrounded by an octahedra whose vertices are occupied by oxygen 
atoms forming $B-$O$-B'$ bonds. Therefore, ordered systems
comprise of alternating corner sharing $B$O$_6$ and $B'$O$_6$ octahedra stacked in all directions. The
$A$-site ions occupy the interstitial voids created by these octahedra.
Flexibility in the choice of these ions provide a large room to obtain suitable combinations for desired properties. For $B$ = Sc$^{3+}$, Y$^{3+}$ 
(trivalent cations) and $B'$ = Nb$^{5+}$, Sb$^{5+}$, Ta$^{5+}$ (pentavalent cations) with either unfilled ($d^0$) or fully-filled ($d^{10}$) 
valence $d$-orbitals are nonmagnetic with large predicted electronic bandgaps \cite{Eng2003, Ray2016, Ray2017, Mondal2018, Himanshu2022}.

The combination $A$ = Ba, Sr, Ca; $B$ = Sc, Y; $B'$ = Nb, Ta, Sb, considered herein, offers a rich platform to systematically explore the role 
of $A$-, $B/B'$-site cations. Their electronic and optical properties have been studied \cite{Eng2003, Ray2016, Ray2017, Mondal2018, Himanshu2022}  and 
those are largely governed by their $B$/$B'$-O bond lengths, $\measuredangle B\text{-O-}B'$ angle, and covalency factors such as the hybridization, 
covalent/ionic radii and electronegativity in a complex way.

The considered compounds belong to two crystal structure types \cite{Faik2012,Ray2016, Ray2017,Mondal2018, Himanshu2022}:
Due to large $A$-site cation, Ba-based compounds, {\it viz.} \bss\ (BSS), \bst\ (BST), \byt\ (BYT), and \byn\ (BYN) crystallize in the face centered cubic 
structure ($Fm\bar{3}m$), featuring linear $B$-O-$B'$ bonds ($\measuredangle B\text{-O-}B'=180^{\circ}$). With decreasing size of the $A$-site cation, 
such as in \sss\ (SSS) and \css\ (CSS), tilting of the $B$O$_6$ and $B'$O$_6$ octahedra (see Fig. \ref{fig:df_bands1}(a)) leads to monoclinic structures 
($P2_{1}/n$) with $c/a=1.01$ and $1.04$, respectively, and $\measuredangle B\text{-O-}B' \neq 180^{\circ}$ \cite{Ray2017, Faik2012}. 
Table SI 
in Supplementary Material (SM) \cite{esi} lists the crystal structure details determined from the XRD data at room temperatures.
The obtained and predicted structures agree with the often-used Goldschmidt tolerance factor \cite{Goldschmidt1926} and are also mostly 
consistent with the recently introduced Bartel's tolerance factor \cite{Bartel2019} (see SM for details).
 
\begin{figure}[b]
	\centering
\includegraphics[scale=0.705,angle=0]{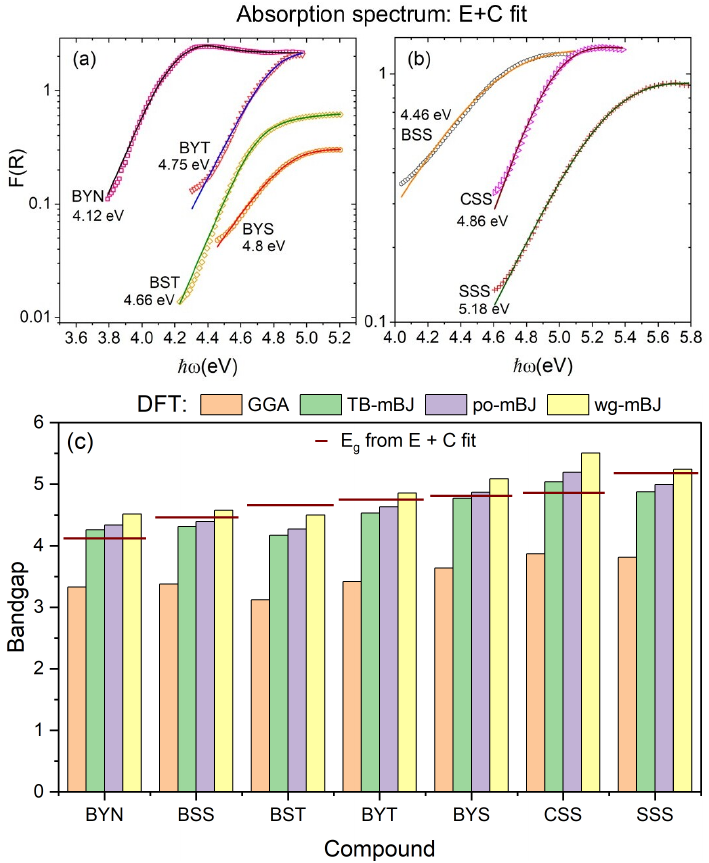}
	\caption{\textbf{Bandgaps. } (a)-(b) Exciton+Continuum (E+C) fit of the of the absorption data in the high-energy range of photon energies (continuum limit), used to obtain the bandgaps. 
	(c) Predicted bandgaps using different flavors DFT functionals/potentials. The brown horizontal lines are experimental values 
	based on the E+C fit.
	}
\label{fig:gaps}
\end{figure}

Optical gaps were determined from the diffuse reflectance spectroscopic data in the UV-Vis range (UV-Vis spectrum) after converting to an effective absorption spectra via the Kubelka-Munk (KM) function $F(R)$ \cite{Davis1970} 
The low-energy linear intercept of $[F(R)\hbar\omega]^n$ as a function of the incident photon energy $\hbar\omega$ provides information about the bandgap. $n=1/2$ and $n=2$, respectively,
corresponds to indirect and direct optical transitions. To account for possible excitons, the optical gaps were also obtained from the effective absorption spectra using the Exciton+Continuum model (E+C model) \cite{Himanshu2022} 
(see Methods for details).
Figure \ref{fig:gaps}(a)-(b) shows the fit of the E+C model to the high energy part of the absorption spectrum for all the considered compounds while Fig. \ref{fig:gaps}(c) shows the values based on density functional (DF) calculations. 
Evidently, all the compounds have 
gaps $4 \text{ eV} < E_g < 5.2\,$eV, 
consistent with previous studies \cite{Eng2003, Ray2017,Himanshu2022}, 
and thus belong to the wide-gap insulators class. It turns out that the E+C model consistently predicts a higher gap values than the KM function (see SM \cite{esi} for comparison of bandgaps from different methods).

Notably, {\it all} the synthesized compounds were off-white in color albeit with differing RGB compositions (different shades). 
This discrepancy suggests possible presence of exciton modes in the visible range. 
Indeed, the log-scale plots in sub-gap region suggests presence of multiple well-defined peaks in the absorption spectra, as discussed later. 

DF calculations, using the all-electron full-potential linear augmented plane waves (FP-LAPW) as implemented in the WIEN2k 
code \cite{wien2k,Blaha2020}, find the ground state to be nonmagnetic and insulating for all the considered compounds, as expected. 
Within GGA, a well-defined bandgap $> 3.0$ eV is found in all cases, as shown in Fig. \ref{fig:gaps}(c), matching with previously reported 
values \cite{Eng2003, Ray2017, Himanshu2022} Changing the $B'$-site cation leads to enhancement of bandgaps with smaller cation size, from 
BYN $\rightarrow$ BYT $\rightarrow$ BYS as well as BST $\rightarrow$ BSS, while changing the $B$-site cation from Y ($4d$) to Sc ($3d$) leads 
to lowering of bandgaps. At the same time, changing the $A$-site cation also increases the bandgaps with smaller ion size.
Importantly, the nature of bandgap is direct (at $\Gamma$) 
in BYN, BYT, BYS, BST and SSS, and indirect in BSS and CSS ($\Gamma \rightarrow X$). 
Even though, the trend in bandgap is consistent with the gaps found using the KM function
(see Methods and Fig \ref{fig:gaps}(c) for details), these values are severely underestimated --- a well-known 
issue with standard DFT \cite{MoralesGarca2017,Perdew2009}.

In order to address the issue of bandgap underestimation in DFT, 
we employ the efficient as well as accurate modified Becke-Johnson (mBJ) potentials  \cite{Singh2010, Kim2010, Martinez2012, Li2013, Borlido2019}. 
Several flavors of the mBJ potentials are now available, such as the Tran-Blaha mBJ (TB-mBJ) \cite{Tran2009}, and customized mBJ potentials suitable for perovskite oxides (`po-mBJ') and 
wide-gap materials (`wg-mBJ') \cite{Koeller2012,Jishi2014}. All the considered mBJ-potentials show significant improvement 
over the GGA bandgap values (see Fig. \ref{fig:gaps}(c)) 
while maintaining the nature of the bandgaps. 
Notably, the choice of DFT parameters, such as $R_{\rm MT}$ influences the electronic and optical properties only quantitatively. For brevity, we will consider the {po-mBJ
for further analysis. 

\begin{figure*}[ht!]
\centering
\includegraphics[width=1.50\columnwidth, angle=0]{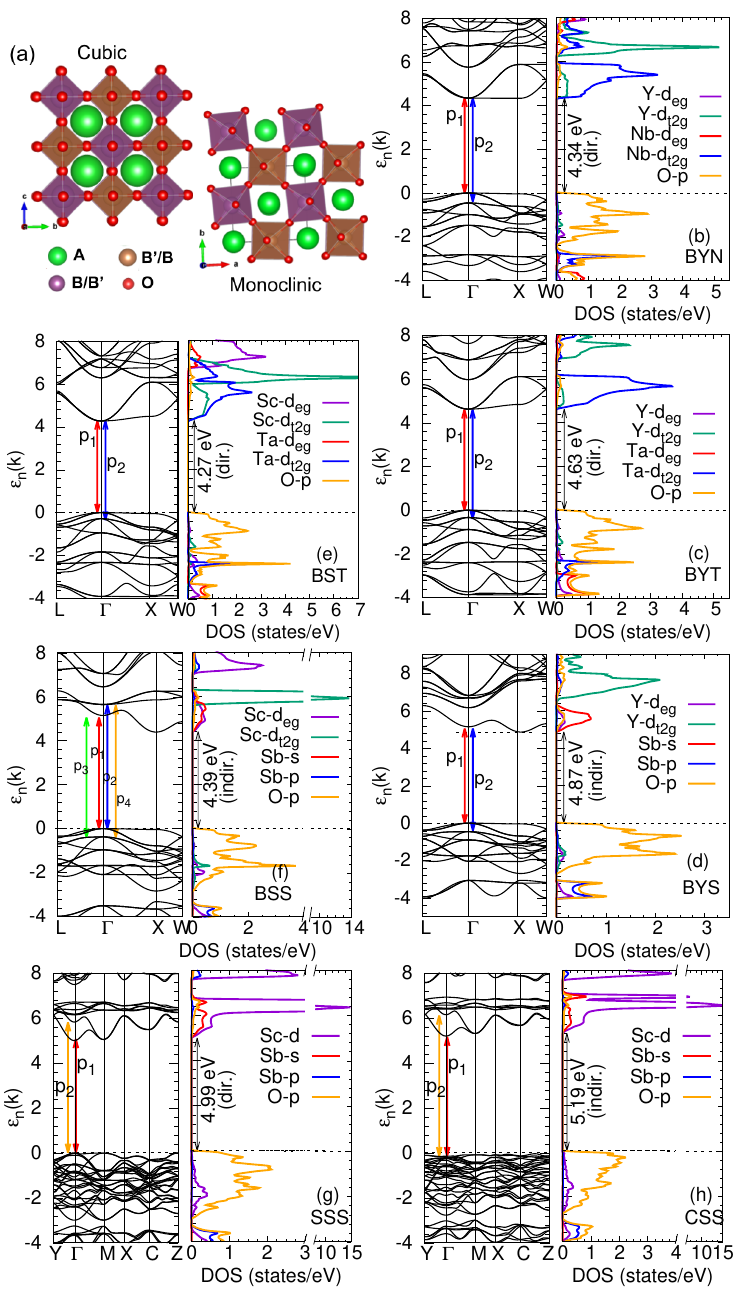}%
	\caption{\textbf{Electronic properties. } (a) Crystal structures of cubic and monoclinic DPOs.
	(b)-(h) Bandstructures and orbital-resolved (partial) densities of states of the considered $d^0$-DPOs obtained using po-mBJ. The arrows indicate the bands considered for detailed study of 
	low-energy dielectric response in Fig. \ref{fig:df_optical}.}
	\label{fig:df_bands1}
\end{figure*}

The electronic bandstructures and atom/orbital-resolved (partial) densities of states are shown 
in Fig. \ref{fig:df_bands1}.  
Consistent with the formal valencies, as well as previous reports, 
the valence band edge is dominantly of O-$2p$ character from the non-bonding oxygen states with negligible traces of other atoms/orbitals.
In the presence of $B{\rm O}_6$ (and $B'{\rm O}_6$) octahedra, the valence $d$ states of the transition metals 
undergo octahedral crystal field splitting, leading to the lower-lying $t_{2g}$ and the higher energy $e_{g}$ states. The conduction 
band edge for compounds not containing Sb is, therefore, dominated by metal $d$-$t_{2g}$ states from either $B$ or $B'$ site or both, which have 
three-fold degeneracy at $\Gamma$ not accounting for spin.
Presence of Sb$^{5+}$ ions, on the other hand, leads to a state with dominant $5s5p$ character at the CB edge. The $B$-site metal $d$-$t_{2g}$ states 
lie next (to the Sb-$s$ state) in the conduction band. Therefore, the low-energy electronic and optical properties are governed by these states 
near the Fermi energy. 

Specifically, the direct gaps in BYN and BYT involve triplet states with dominant Nb-$t_{2g}$ or Ta-$t_{2g}$ characters, 3-fold degenerate at 
$\Gamma$, at the CB edge. Small contributions from Y-$t_{2g}$ states at the CB edge, $\sim 10\%$ for BYN and $\sim 6\%$ for BYT is also noted. 
In comparison, the larger but indirect gap in BYS is between the O-$2p$ dominated states in valence band (VB) and the non-degenerate Sb-$5s5p$ state 
at the CB edge, from $\Gamma - X$ (see Fig. \ref{fig:df_bands1}(d)). At the same time, the relative Y-$d$ contribution at the CB edge is much larger. 

Presence of Sc at the $B$-sites in BST leads to a significantly larger contribution ($\sim 20\%$) by the Sc $d$-$t_{2g}$ orbitals along with 
Ta $d$-$t_{2g}$ orbitals due to smaller radial extent of the valence Sc-$3d$ states compared to Y-$4d$ states, leading to large overlap with the 
O-$2p$ states \cite{Eng2003}. In BSS, the relative gap between the Sb-$5s5p$ and Sc $t_{2g}$ states is much smaller than in BST while the Sc 
contribution at the CB edge is comparable to that of Sb; nevertheless the CB edge is formed by the Sb-$5s5p$ states. 

On the other hand, changing the $A$-site cation leads to symmetry lowering of the crystal structure, which in turn significantly alters the 
bandgaps even if their contributions at the band edge is tiny. The bandgap increases by $\sim$0.6 and $0.8$ eV, respectively for SSS and CSS 
(see Fig. \ref{fig:gaps}(c)) \cite{Eng2003, Ray2017}. However, as shown in Fig. \ref{fig:df_bands1}, in all the Sb based systems, Sb-$5s5p$ states 
contribute dominantly at the CB edge. Deviation from cubic structure enhances the gap, but only quantitatively influences the band edges and related 
optical properties \cite{Ray2017}.

Essentially, all the synthesized compounds are oxygen to transition metal charge transfer complexes 
with large electronic bandgaps. In all cases, the VB edge is formed by the O-$2p$ states whereas an isolated set of bands lie at the CB edge. 
Details of the CB edge depends on the composition: Variations in $B$/$B'$-site ions can lead not only to transition from direct to indirect gaps, 
but also change the nature of the CB edge states (transition metal $d$ states vs Sb-$5s5p$ states). At the same time, quantitative differences 
in terms of atom/orbital contributions exist. Of particular relevance are the dispersionless bands along $\Gamma - X$ at the band edges as well as 
presence of linear dispersions around the $W$-point in non-Sb compounds. As they directly influence the effective mass of the electrons and holes, 
and the optical transition matrix elements \cite{Dresselhaus2018}, these features may have interesting consequence for the excitons in these systems. 

\subsection{Exciton modes \& Optical properties}

We now turn our attention to the primary focus of this study --- presence of excitons and their possible correlation(s) with the electronic 
structure. In principle, as in double perovskites halides, both hydrogenic (Wannier-Mott) and non-hydrogenic excitons could be present \cite{Biega2023}. 
Moreover, the present exciton modes may display a wide range of localization behavior as in transition metal oxides and perovskite 
oxides \cite{Lany2015, Biswas2018, Varrassi2021}.

To identify the exciton modes, we analyze 
the observed low-energy absorption data and fit the low-energy peaks using the Elliot formula \cite{Elliott1957, Wang2017}, as discussed 
in detail in Ref.~\cite{Himanshu2022} (also see Methods). 
Although the Elliot formula used here relies on hydrogenic model of excitons, the procedure utilized here merely identifies the exciton peaks and 
the corresponding exciton binding energies and the electronic gap (presumed optical gap without excitons) of the materials. The hydrogenic model 
suggests that the exciton peaks scale as $E_n = -(\mu/\epsilon_{\infty})\,1/n^2$, where $\mu$ is the effective mass 
and $\epsilon_{\infty}$ is the static dielectric constant. In order to characterize the nature of excitons as non-hydrogenic, in principle, detailed 
knowledge of dielectric screening is required which is, however, not available for the considered compounds at present. Nevertheless, 
deviations from hydrogenic model can be obtained through a comparative analysis of the peak positions and their shapes.
It is important to note that, since the absorption data for an indirect bandgap in presence of excitons mimic that of a direct bandgap system, the 
KM function 
is obtained assuming all materials to possess direct bandgap \cite{Yu_book2010}. 

\begin{figure}[ht!]
	\centering
\includegraphics[width=1.00\columnwidth,angle=0]{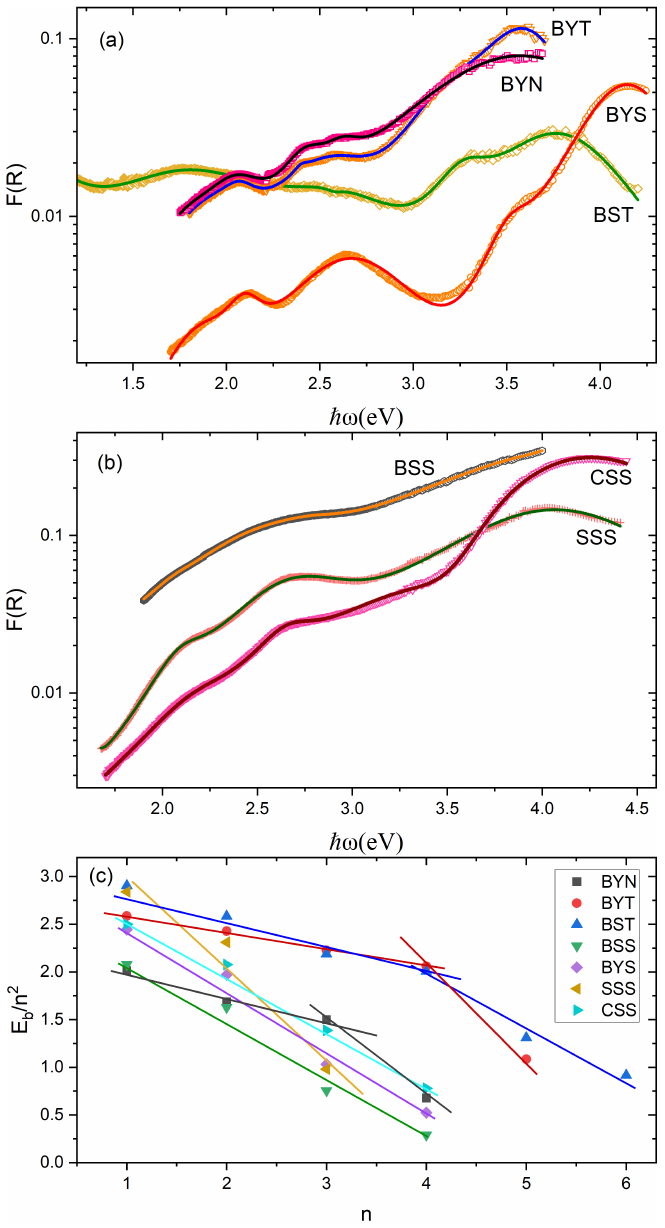}
	\caption{\textbf{Exciton characteristics. } (a)-(b) Low energy part of the absorption spectrum (symbols). The corresponding multi-peak fits assuming hydrogenic model for excitons 
	is shown by solid lines. 
(c) $E_b/n^2$ vs $n$ for the low-energy peaks identified from the data, showing significant deviation from the Wannier Mott excitons. 
	Solid lines are guide to the eye.}
\label{fig:exciton_fits}
\end{figure}

Figures \ref{fig:exciton_fits}(a)-(b) show the low-energy absorption spectra and the multiple-peak fit to the data.  
The details of the fit quality are provided in Table SII \cite{esi}.  
Evidently, in all the compounds considered, low-energy peaks are present well below the electronic gaps. 
Remarkably, these peaks extend well within the visible range of the electromagnetic spectrum (1.65 eV - 3.25 eV). The corresponding binding energies 
are found to lie between 2 eV to 3 eV. However, the peak intensities are quite small, leading to weak absorption in the visible range. This explains 
the origin of off-white colors of the considered compounds. 

For cubic compounds not involving Sb (BYN, BYT, and BST), {\it i.e.} for compounds with $d$-$t_{2g}$ states at CB edge, one expects them to 
be well-described by the hydrogenic model as the systems are not only isotropic due to the cubic symmetry, but also 
possess direct bandgaps \cite{Biega2023}. Additionally, the exciton characteristics is likely to be governed by the joint density of 
states (JDOS) for transitions from O-$2p$ states to $d$-$t_{2g}$ states, as in perovskite oxides \cite{Varrassi2021}. 
The obtained $E_b/n^2$ vs. $n$ in Fig \ref{fig:exciton_fits}(c), however, shows a weak deviation from the ideal $E_n \propto 1/n^2$ behavior 
for small $n$. Interestingly, large deviation is obtained for large $n$, suggesting different possible origins of these peaks. 

In comparison, the Sb compounds show significantly large deviations for all values of $n$ presumably due to presence of indirect bandgaps leading to 
non-hydrogenic excitons in all cases. (In Ca, the energy at the VB edge at $M$ is $<2$ meV below the Fermi energy, $\epsilon_{\rm F}$.) The 
binding energies are smaller than others. Smaller binding energies arise due to weaker Coulomb interaction between the electrons and holes forming 
the excitons. Contradictory to that, the screening in Sb-compounds is much weaker as evidenced by their static dielectric constant values, albeit 
without many body effects, as shown in Fig. \ref{fig:df_optical}(a).
Therefore, it seems that the large radial extent of the involved $5s5p$ state compete with the screening effect, eventually leading to weak 
interaction between the electron-hole pair. 

Role of indirect bandgaps in the nature of excitons in Sb-compounds is further attested by the fact that linewidths of the peaks are much larger 
than in others (see Table \ref{tab:summary} and SI \cite{esi}) to the extent that low-energy absorption spectrum in BSS is rather featureless. At the same 
time, the absorption in ScSb-compounds is also higher.
To gain further insights, we study the band-resolved imaginary part of the dielectric response $\sum_{\{n\},\{m\}}\epsilon_2(\{n\},\{m\},\omega)$, 
involving band sets $\{n\}$ and $\{m\}$ across $\epsilon_{\rm F}$. The 3-fold degenerate $t_{2g}$ bands ($\{m\}=\{N+1, N+2, N+3\}$) at the CB 
edge and the sets of O-$2p$-dominated bands degenerate at $\Gamma$, $\{n\}=\{N, N-1, N-2\}$ and/or $\{n\}=\{N-3, N-4, N-5\}$ contribute at 
low-energies, $N$ being the number of electrons in the system. Sum over bands lead to $\epsilon_2^{p_1}(\omega)$ and $\epsilon_2^{p_2}(\omega)$, 
respectively, which contribute at low energies.

\begin{figure*}[ht!]
	\centering
\includegraphics[width=1.510\columnwidth,angle=0]{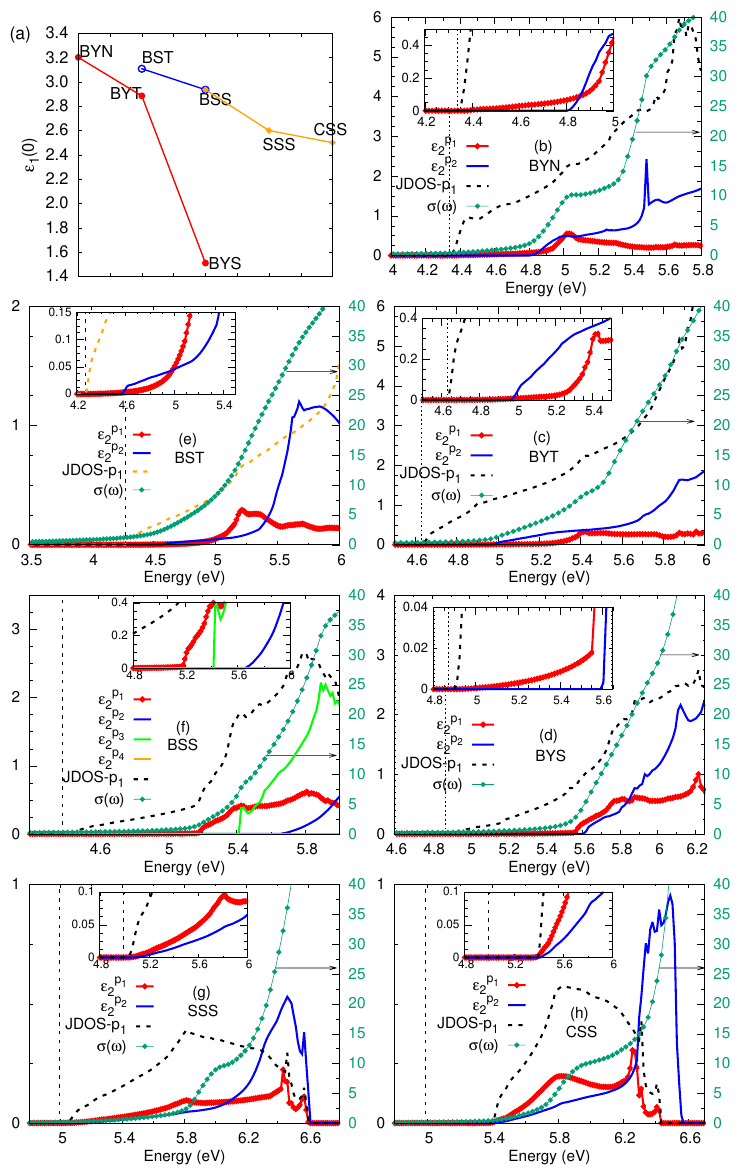}
	\caption{\textbf{Optical properties. } (a) Static dielectric constant, and (b)-(h) band-resolved dielectric responses, joint density of states (JDOS) for $p_1$ contributions (see Fig. \ref{fig:df_bands1}) 
	and the total optical 
	conductivity $\sigma(\omega)$. Note that for CSS and SSS, the dielectric constant is averaged over all the components ($x$,$y$,$z$ directions), whereas the $p_1$ contributions 
	are summed over.}
\label{fig:df_optical}
\end{figure*}

Figure \ref{fig:df_optical} shows the low-energy contributions to the dielectric function and the related JDOS.
In all cases, vanishing optical-transition matrix elements despite large JDOS lead to sizable response away from the bandgap energies, implying 
no contribution at the $\Gamma$-point, contrary to perovskite oxides \cite{Varrassi2021}. 
Nevertheless, $\gtrsim 1$ eV widths of the dominant peaks in the band-resolved dielectric function (see SM for details) suggest rather finite 
extent of the identified excitons involving these edge states. 
Furthermore, the relatively smaller contribution of the edge transitions ($p_1$) compared to other contributions such as the sub-edge 
transitions $p_2$ may explain the weak intensity peaks in the absorption data. In fact, in some cases, the $p_2$-contribution even starts at 
lower energies as shown in the insets of Fig. \ref{fig:df_optical}.

BYN and BYT display large edge contributions arising from optical transitions between the sub-edge O-$2p$ below $\epsilon_{\rm F}$ and the $t_{2g}$ states 
forming the $p_2$ contribution.
At the same time, changing the $B'$ site, from Nb to Ta, results in no $p_1$ contribution at the (response) onset. The band-resolved 
contributions \cite{esi} further suggests no $p_1$ contributions from the dispersionless (flat) band along $\Gamma - X$ in both cases. In fact, in BYN, 
dominant contribution to arises close to the $W$- and $X$ points whereas the gradual rise is likely due to contributions along $\Gamma-L$. In 
comparison, $k$-points along $\Gamma$-$X$ contribute to the onset in BYT through $p_2$. 
It is interesting to note the relative contribution of Y-$d$-$t_{2g}$ states at the CB edge is also slightly larger in BYN due to larger radial extent 
of $5d$ states in Ta. 

Presence of Sc at the $B$-site, in BST, despite significant Sc contribution at the band edge (Fig. \ref{fig:df_bands1}), also leads to the response 
onset due to $p_2$ contributions only. While the band-resolved contributions (see SM) is similar to BYT, $p_1$ contributions, however, show a 
dominant peak at low energies $\sim$5.5 eV, distinct from both BYT and BYN. In terms of the linewidths (FWHM), ($\Gamma$ in Eq. (\ref{eqn:elliot})), 
the lowest energy peak in these three compounds suggests similar radial extent as well as lifetimes of the excitons, likely of similar origin. 
Other peak characteristics are also remarkably similar between BYN and BST. The details of the peak fit and their characteristics are presented in 
Table SII \cite{esi}. 

In sharp contrast, the singular $5s5p$ contribution at the CB edge in Sb-compounds with cubic symmetry leads to rather sharp response onset 
characterized by $p_1$ contribution as compared to the non-Sb cubic compounds.
Detailed analysis of the band-resolved contributions suggests non-vanishing contribution from the $\Gamma$-point unlike the non-Sb 
counterparts along with  sizable contributions at the $W$-point. Remarkably, the well separated Sb state at the CB edge leads to 
comparable $p_1$ and $p_2$ contributions, unlike in BSS where, despite the "hybridization" of the Sc-$d$-$t_{2g}$ states with the Sb state 
at the CB edge, only $p_1$ contributes to the response onset.  

The observed low-energy spectra in BSS show a rather featureless absorption (much larger linewidths). Delocalized valence $5s$ and $5p$ states, 
compared to $d$ states, together with exciton-phonon interactions at the indirect gap edge leads to manifold enhancement in the  broadening ($\Gamma$) 
of the lowest exciton peaks for BYS as well as BSS.
Other higher lying exciton peak-width are also influenced by presence of Y/Sc ions, suggesting different origin and nature of the observed exciton peaks.

\section{Conclusions and Outlook}

To summarize, we find existence of multiple low-intensity but well-defined low-energy peaks in the absorption spectra of the considered $d^0$-DPOs with 
large electronic gaps $>4$ eV. These peaks have been identified as the exciton peaks and the corresponding binding energies were found to be as large 
as $\sim 3.0$ eV, extending into the visible range in all cases (see Table \ref{tab:summary}). Presence of such excitons naturally explains the observed off-white colors of the samples. 

Comparison with the DF-based electronic and optical properties suggests that low intensities of the observed exciton peaks originates from the small 
optical matrix elements between the band edge states across the Fermi energy, and in many cases, the sub-edge optical transitions are dominant. 
These findings are largely independent of the choice of the functional used in DFT. Our calculations further suggest that these systems are well-described by 
both po-mBJ and wg-mBJ, suitable for perovskite oxides and wide-gap insulators, respectively.

\begin{table}[ht!]
	\caption{\label{tab:summary} {\bf Summary of exciton properties. } Electronic bandgaps, $E_g$, obtained from E+C fits, binding energy, $E_b$, and the 
	linewidth, $\Gamma$, of the first exciton peak ($n=1$). Band-edge contributions are also provided. For $E_g$, nature of the gap 
	ascertained from DFT is mentioned in brackets.}
    \begin{tabularx}{0.48\textwidth}{ p{1.0cm} p{1.8cm} p{1.15cm} p{1.15cm} p{1.85cm} }
    \hline \hline
        Comp. & $E_g$ (eV) & $E_b$ (eV) & $\Gamma$ (eV) & Dominant contrib. \\
        \hline \hline
        BYN & 4.12 (dir.) & 2.01 & 0.14 & $p_1$, $p_2$\\
        BYT & 4.75 (dir.) & 2.59 & 0.07 & $p_2$\\
        BYS & 4.80 (indir.) & 2.44 & 0.08 & $p_1$\\
        \\
        BST & 4.66 (dir.) & 2.91 & 0.39 & $p_2$\\
        BSS & 4.46 (indir.) & 2.08 & 0.34 & $p_1$\\
        \\
        SSS & 5.18 (dir.) & 2.84 & 0.12  & $p_1$, $p_2$\\
        CSS & 4.86 (indir.) & 2.50 & 0.20  & $p_1$  \\    
    \hline
    \end{tabularx}
\end{table}

Importantly, the binding energies of these excitons show deviations from the $\sim 1/n^2$ dependence for the hydrogenic model. 
In comparison to cases where the CB edge is formed by localized $d$ states, presence of delocalized $5s5p$ states in the Sb-compounds and 
presence of indirect bandgaps lead to considerably larger deviations from the hydrogenic models and large exciton linewidths. 
Similar behavior is expected for the recently reported iodine based $d^0$-DPO \cite{Jiang2022,Ali2024}, where the CB edge is formed by the I-$5s5p$ states.
At the same time, in Sb-compounds, the binding energy is found to be surprisingly smaller along with weak dielectric screening. From our analysis, 
however, the relative contribution of the delocalized nature of the edge states and the role of indirect bandgaps to this effect are not discernible. 
Therefore, full scale DF calculations including the many body effects should be of interest.

Large binding energies of these excitons, in principle, render them interesting from the potential optoelectronic applications viewpoint. However, 
the intensities in the absorption spectra are low.
Our analysis highlights the complex relationship between electronic structure and the nature of excitons. Nonetheless, in principle, 
reliable information about the nature of excitons can be obtained from computationally cheap DF calculations. This bolsters the notion that machine learning based 
approaches combining standard DF electronic structure calculations can yield quantitative prediction for materials. 

\begin{appendix}
\section{\label{sec:methods}Methods}

\subsection{Experimental Methods}

The conventional solid state reaction method was used to create polycrystalline samples of Ba$_2$YB'O$_6$ (B' = Nb, Ta, Sb), by heating at 
1400 $^{\circ}$C for 12 hours, while Ba$_2$ScSbO$_6$ was prepared under the same conditions for 72 hours. The interacting substances, 
BaCO$_3$ (99.999\%), Y$_2$O$_3$ (99.99\%), Sc$_2$O$_3$ (99.99\%) and Ta$_2$O$_5$, Sb$_2$O$_5$ and Nb$_2$O$_5$ (99.99\%) from Sigma-Aldrich, 
were combined, crushed in an agate mortar with acetone, and then heated in crucibles. The final samples are cooled to room temperature and 
ground for characterization.
Polycrystalline A$_2$ScSbO$_6$ (A = Sr, Ca)  were prepared following methods in Ref. \cite{Faik2012} while details of and Ba$_2$ScTaO$_6$ 
can be found in Ref. \cite{Himanshu2022}. 

All the synthesized samples were found to be off-white in color. X-ray powder diffraction (XRD) was used to record the sample phase using Bruker D8 Advance X-ray diffractometer. 
The resulting XRD patterns and the corresponding crystal structure parameters are provided in SM \cite{esi}.
The diffuse reflectance spectrum were obtained using Perkin-Elmer Lamda 950 UV/Vis/NIR spectrophotometer. 

\subsection{Density functional (DF) calculations}
\label{sec:methods-dft}

DF calculations were performed using the all electron full-potential linearized augmented plane wave (FP-LAPW) method
within the scalar relativistic approximation, as implemented in the WIEN2k
code \cite{wien2k,Blaha2020}. The experimental crystal structures were considered (see Table SI \cite{esi} for details). 
A $18\times18\times18$ ($14\times10\times14$) $k$-mesh in the full Brillouin zone (BZ) were considered for the cubic (monoclinic) compounds in order to carry out
the integrals over the BZ. The muffin-tin radii for Ba, Sr, Ca, Sc, Y, Nb, Ta, Sb and O
were kept fixed at 2.5 a.u., 2.31 a.u., 2.19 a.u., 2.05 a.u., 2.14 a.u., 1.89 a.u., 2.04 a.u., 2.04 a.u. and 1.71 a.u.,
respectively. The presented results correspond to $R_{MT} \times k_{\rm max} =8$ for all cases, where $k_{\rm max}$ is the plane-wave 
cut-off and $R_{MT}$ is the smallest muffin-tin radii among all atoms. The exchange and correlation
effects have been treated within the generalized gradient approximation (Perdew-Burke-Erzenhof (PBE)
implementation \cite{Perdew1996}). Self-consistent solutions correspond to convergence
below $10^{-5}$ e/a.u.$^3$ for charge, and $10^{-6}$ Ry for the total energy per unit cell. Variations in $R_{MT}$ values affects the 
total energies but shows only quantitative effects on the electronic properties. Changes in bandgap were typically less than 0.01 eV. 
Similarly, spin-orbit effects were also considered in some cases but were found to have negligible effect on the bandgaps and optical properties.

To address the issue of bandgap underestimation
with standard DFT functional \cite{Baerends2013}, we also employed different available flavors of the modified Becke-Johnson (mBJ)
exchange-correlation potentials \cite{Tran2009, Koeller2012,Koeller2012, Jishi2014}.
The complex dielectric function \cite{Draxl2006}
\begin{equation}
    \varepsilon(\omega) = \varepsilon_1(\omega) +i \varepsilon_2(\omega)\,
\end{equation}
were calculated to study the optical properties.
In general, $\varepsilon \equiv \varepsilon_{\mu \nu}$ ($\mu,\nu = x,y,z$), $\varepsilon_1$ and $\varepsilon_2$ are second rank tensors with nine components. However, depending on the crystalline symmetry, only a few of 
these components could be independent. For a cubic structure which is also isotropic, for example, the three principal directions ($x$, $y$, and $z$) are equivalent, leading 
to only one independent value. 
To further analyze the low-energy contributions to the optical response, band-resolved contributions were studied. Contributions from degenerate 
bands at $\Gamma$ were summed over, as indicated in Fig. \ref{fig:exciton_fits}. Specifically, in cubic systems, contributions from the 3-fold 
degenerate $t_{2g}$ bands ({$N + 1$, $N + 2$, $N + 3$}) at
the CB edge and the sets of O-$p$ dominated bands, degenerate at $\Gamma$ ({$N$, $N-1$, $N-2$} and/or {$N-3$, $N-4$, $N-5$}) were considered, 
where $N$ is the number of electrons in the system.

\subsection{Fitting of the UV-Vis spectrum}

The electronic gap was determined using the Kubelka-Munk (KM) function, $F(R)$ \cite{Davis1970, Barton1999} obtained by converting the reflectance 
spectrum into an effective absorption spectrum \cite{Ray2017}.
\begin{equation}
  F(R) \propto \frac{(\hbar\omega -E_g)^n}{\hbar\omega} \,, 
\label{eqn:km}
\end{equation}
where $n=\frac{1}{2}$ for direct-allowed transitions. $E_g$ is the bandgap and $\hbar\omega$ is the incident photon energy. The bandgap is 
computed by locating the intercept of the linear section of the 
$[F(R) \hbar\omega]^2$ vs
$\hbar\omega$ curve on the energy axis \cite{Ray2017}. 

As in the case of {\bst} \cite{Himanshu2022}, multiple low-intensity peaks appear in the absorption spectrum (KM function, 
$F(R)$) within the energy range of $\sim 1.5$ eV to $\sim 4.5$ eV, as shown in the log-scale plot in Fig. \ref{fig:exciton_fits}(a). 
Typically, large (unphysical) fluctuations were observed in the infrared region, which were discarded from all the data when modelling the low-energy
part of the data. The peaks in 
Fig. \ref{fig:exciton_fits}(a) correspond to exciton modes. To analyze them, the absorption data was modeled using the Elliott formula for 
direct-gap excitons \cite{Elliott1957,Wang2017}, as detailed in \cite{Himanshu2022}. To summarize, according to this model, absorption is expressed as 
a combination of excitonic contributions, 
$\alpha_{nx}$ at lower photon energies (below band edge) and a continuum, 
$\alpha_{\rm cont}$ at higher photon energies. The total contribution to the absorption coefficient is thus given by \cite{Himanshu2022}:
\begin{eqnarray}
	\alpha(\omega) &=& \sum_n \alpha_{nx}(\omega) + \alpha_{\rm cont}(\omega) \nonumber \\
	&=& \sqrt{E_b}\,\, \Big[ \sum_n 2A_n \,\, \frac{E_b}{n^3}\,\,  {\rm sech} \Big( \frac{\hbar\omega - E_g + E_b/n^2}{\Gamma_n} \Big) \nonumber \\
	&+& B \int_{E_g}^{\infty} {\rm sech} \Big( \frac{\hbar \omega - E}{\Gamma^{'}} \Big) \cdot \nonumber \\ 
	& \quad& \frac{1 + 10\frac{m^2}{\hbar^4}EC_{np} +
                    \big( \frac{\sqrt{126}m^2}{\hbar^4}EC_{np} \big)^2}{1 - \exp{\big( -2\pi \sqrt{\frac{E_b}{E-E_g}}\big) } }\, dE \,\, \Big] \,.
\label{eqn:elliot}
\end{eqnarray}
Here, the parameter $n$ denotes the order of the exciton state, with $\alpha_{nx}$ representing the absorption corresponding to the 
$n$-th exciton state. The `sech' function serves as a broadening function for the exciton lineshape, characterized by a linewidth 
$\Gamma_{n}$. Here, $E_g$ represents the electronic bandgap, while $E_b$ denotes the exciton binding energy. The correction factor 
for deviations from parabolic bands is given by $C_{np}$. The parameters $B$ and $\Gamma'$ correspond to the amplitude and broadening of the 
continuum part, respectively. Additionally, $m$ represents the free electron mass, and $\hbar$ is the reduced Planck constant.

A least-squares fit of the expression in Eq. (\ref{eqn:elliot}) to the data was performed using the \textit{python-lmfit} function along 
with SciPy. For high-energy absorption data fitting, Origin software was used. To simplify the analysis, the low-energy region containing 
the exciton peaks (marked by arrows in Fig. \ref{fig:exciton_fits}(a)), and the high-energy region were treated separately.

In the high-energy range, deviations between the calculated and observed spectra were noted for photon energies above $\hbar\omega^{\rm{cutoff}}$ 
(e.g., $\sim$5.2 eV for {\bst}; see Fig. \ref{fig:gaps}(a)-(b)). As a result, the fitting was limited to a maximum energy of $\hbar\omega^{\rm{cutoff}}$. 

Different models were explored for fitting the high-energy region of the absorption data. First, considering the exponential onset of the 
absorption spectrum in this energy range, the data was modeled using a Urbach tail 
[$U(\omega) = U_0 \,\exp\{(\omega - E_1)/E_{\rm U}\}$, where $E_{\rm U}$ represents the Urbach energy, and $E_1$ and $U_0$ are 
fitting parameters. The obtained values of the Urbach Energies were: $E_{\rm U} = 0.12$ eV for BYN, 0.14 eV for BYT, 0.12 eV for BST, 0.23 eV for BYS, 
0.55 eV for BYS, 0.37 eV for SSS, and 0.30 for CSS. 

Second, a model 
with $\alpha(\omega)$ for a single exciton mode with a suitable value of $n$ yielded the best-fit parameters. The fit to the data is 
shown in Fig. \ref{fig:gaps}(a-b).
For the low-energy portion of the data, a fit of \( \alpha^{\prime}(\omega) = \sum_n \alpha_{nx} \) was successfully obtained, with $n$ exciton 
mode peaks. It is important to note that the data near the first peak is asymmetric for all the compounds under consideration. To accurately 
determine the peak position, a hypothetical peak at the edge of the data was considered to account for the offset in the first peak. As a result, 
a high quality fit to the data was obtained, shown in Fig. \ref{fig:exciton_fits}(d). 
Table SII \cite{esi} lists the optimized fit parameters for all the compounds considered in the study. These parameters include the exciton 
binding energy \( E_b \) and the peak linewidth \( \Gamma_n \) for each exciton mode, indicating the variation in exciton behavior across the compounds.

The electronic gaps obtained from the high energy part and the low-energy part of the UV-Vis data are in good agreement with each other, typically 
within 5\%. In all cases, the reported electronic gap from the data corresponds to the values obtained from the E+C fits. 

\end{appendix}

\section*{Supplementary Material}
The Supplementary Material provides details of the crystal structure, XRD pattern, methods, exciton characteristics, and further details of the electronic and optical properties.

\section*{Data Availability Statement}
The data that support the findings of this study are available from the corresponding author upon reasonable request.

\begin{acknowledgments}
BB thanks the Director, Variable Energy Cyclotron Center hospitality and access to experimental facilities. BB and AKH thank Director, Variable Energy Cyclotron Center for support. 
We thank Kumar Brajesh and Bibek Kumar Sonu for technical assistance. We also thank Mr. Sovam Maiti for technical assicatance with the HPC facilty at BIT Mesra.
\end{acknowledgments}

%

\begin{thebibliography}{62}
\expandafter\ifx\csname natexlab\endcsname\relax\def\natexlab#1{#1}\fi
\providecommand{\bibinfo}[2]{#2}
\ifx\xfnm\relax \def\xfnm[#1]{\unskip,\space#1}\fi
\bibitem[{Eng et~al.(2003)Eng, Barnes, Auer, and Woodward}]{Eng2003}
\bibinfo{author}{H.~W. Eng}, \bibinfo{author}{P.~W. Barnes},
  \bibinfo{author}{B.~M. Auer}, \bibinfo{author}{P.~M. Woodward},
  \bibinfo{journal}{Journal of Solid State Chemistry} \bibinfo{volume}{175}
  (\bibinfo{year}{2003}) \bibinfo{pages}{94--109}.
\bibitem[{Vasala and Karppinen(2015)}]{Vasala2015}
\bibinfo{author}{S.~Vasala}, \bibinfo{author}{M.~Karppinen},
  \bibinfo{journal}{Progress in Solid State Chemistry} \bibinfo{volume}{43}
  (\bibinfo{year}{2015}) \bibinfo{pages}{1--36}.
\bibitem[{Ray et~al.(2017)Ray, Himanshu, Sen, Kumar, Richter, and
  Sinha}]{Ray2017}
\bibinfo{author}{R.~Ray}, \bibinfo{author}{A.~Himanshu},
  \bibinfo{author}{P.~Sen}, \bibinfo{author}{U.~Kumar},
  \bibinfo{author}{M.~Richter}, \bibinfo{author}{T.~Sinha},
  \bibinfo{journal}{Journal of Alloys and Compounds} \bibinfo{volume}{705}
  (\bibinfo{year}{2017}) \bibinfo{pages}{497--506}.
\bibitem[{Yin et~al.(2019)Yin, Weng, Ge, Sun, Li, and Yan}]{Yin2019}
\bibinfo{author}{W.-J. Yin}, \bibinfo{author}{B.~Weng},
  \bibinfo{author}{J.~Ge}, \bibinfo{author}{Q.~Sun}, \bibinfo{author}{Z.~Li},
  \bibinfo{author}{Y.~Yan}, \bibinfo{journal}{Energy {\&} Environmental
  Science} \bibinfo{volume}{12} (\bibinfo{year}{2019})
  \bibinfo{pages}{442–462}.
\bibitem[{Nair et~al.(2022)Nair, Krishnia, Trukhanov, Thakur, and
  Thakur}]{Nair2022}
\bibinfo{author}{S.~S. Nair}, \bibinfo{author}{L.~Krishnia},
  \bibinfo{author}{A.~Trukhanov}, \bibinfo{author}{P.~Thakur},
  \bibinfo{author}{A.~Thakur}, \bibinfo{journal}{Ceramics International}
  \bibinfo{volume}{48} (\bibinfo{year}{2022}) \bibinfo{pages}{34128–34147}.
\bibitem[{Ji et~al.(2023)Ji, Boschloo, Wang, and Gao}]{Ji2023}
\bibinfo{author}{F.~Ji}, \bibinfo{author}{G.~Boschloo},
  \bibinfo{author}{F.~Wang}, \bibinfo{author}{F.~Gao}, \bibinfo{journal}{Solar
  RRL} \bibinfo{volume}{7} (\bibinfo{year}{2023}) \bibinfo{pages}{2201112}.
\bibitem[{Peters et~al.(2022)Peters, Liu, De~Siena, Kanatzidis, and
  Wessels}]{Peters2022}
\bibinfo{author}{J.~A. Peters}, \bibinfo{author}{Z.~Liu},
  \bibinfo{author}{M.~C. De~Siena}, \bibinfo{author}{M.~G. Kanatzidis},
  \bibinfo{author}{B.~W. Wessels}, \bibinfo{journal}{Journal of Luminescence}
  \bibinfo{volume}{243} (\bibinfo{year}{2022}) \bibinfo{pages}{118661}.
\bibitem[{Wang et~al.(2024)Wang, Liu, Wan, Gao, Wang, Liu, Tan, Guo, Zhao, Hu,
  Li, and Yang}]{Wang2024}
\bibinfo{author}{L.~Wang}, \bibinfo{author}{X.~Liu}, \bibinfo{author}{L.~Wan},
  \bibinfo{author}{Y.~Gao}, \bibinfo{author}{X.~Wang},
  \bibinfo{author}{J.~Liu}, \bibinfo{author}{S.~Tan}, \bibinfo{author}{Q.~Guo},
  \bibinfo{author}{W.~Zhao}, \bibinfo{author}{W.~Hu}, \bibinfo{author}{Q.~Li},
  \bibinfo{author}{J.~Yang}, \bibinfo{journal}{The Journal of Physical
  Chemistry Letters} \bibinfo{volume}{15} (\bibinfo{year}{2024})
  \bibinfo{pages}{2096–2104}.
\bibitem[{Jang and Mennucci(2018)}]{Jang2018}
\bibinfo{author}{S.~J. Jang}, \bibinfo{author}{B.~Mennucci},
  \bibinfo{journal}{Rev. Mod. Phys.} \bibinfo{volume}{90}
  (\bibinfo{year}{2018}) \bibinfo{pages}{035003}.
\bibitem[{Hu et~al.(2023)Hu, Lin, Lynch, Xu, and Jariwala}]{Hu2023}
\bibinfo{author}{Z.~Hu}, \bibinfo{author}{D.~Lin}, \bibinfo{author}{J.~Lynch},
  \bibinfo{author}{K.~Xu}, \bibinfo{author}{D.~Jariwala},
  \bibinfo{journal}{Device} \bibinfo{volume}{1} (\bibinfo{year}{2023})
  \bibinfo{pages}{100003}.
\bibitem[{Datta et~al.(2022)Datta, Lyu, Li, Taniguchi, Watanabe, and
  Deotare}]{Datta2022}
\bibinfo{author}{K.~Datta}, \bibinfo{author}{Z.~Lyu}, \bibinfo{author}{Z.~Li},
  \bibinfo{author}{T.~Taniguchi}, \bibinfo{author}{K.~Watanabe},
  \bibinfo{author}{P.~B. Deotare}, \bibinfo{journal}{Nature Photonics}
  \bibinfo{volume}{16} (\bibinfo{year}{2022}) \bibinfo{pages}{242–247}.
\bibitem[{Biega et~al.(2023)Biega, Chen, Filip, and Leppert}]{Biega2023}
\bibinfo{author}{R.-I. Biega}, \bibinfo{author}{Y.~Chen},
  \bibinfo{author}{M.~R. Filip}, \bibinfo{author}{L.~Leppert},
  \bibinfo{journal}{Nano Letters} \bibinfo{volume}{23} (\bibinfo{year}{2023})
  \bibinfo{pages}{8155–8161}.
\bibitem[{Omelchenko et~al.(2017)Omelchenko, Tolstova, Atwater, and
  Lewis}]{Omelchenko2017}
\bibinfo{author}{S.~T. Omelchenko}, \bibinfo{author}{Y.~Tolstova},
  \bibinfo{author}{H.~A. Atwater}, \bibinfo{author}{N.~S. Lewis},
  \bibinfo{journal}{ACS Energy Letters} \bibinfo{volume}{2}
  (\bibinfo{year}{2017}) \bibinfo{pages}{431–437}.
\bibitem[{Lushchik et~al.(2000)Lushchik, Kirm, Lushchik, Martinson, and
  Zimmerer}]{Lushchik2000}
\bibinfo{author}{A.~Lushchik}, \bibinfo{author}{M.~Kirm},
  \bibinfo{author}{C.~Lushchik}, \bibinfo{author}{I.~Martinson},
  \bibinfo{author}{G.~Zimmerer}, \bibinfo{journal}{Journal of Luminescence}
  \bibinfo{volume}{87–89} (\bibinfo{year}{2000}) \bibinfo{pages}{232–234}.
\bibitem[{Koch et~al.(2006)Koch, Kira, Khitrova, and Gibbs}]{Koch2006}
\bibinfo{author}{S.~W. Koch}, \bibinfo{author}{M.~Kira},
  \bibinfo{author}{G.~Khitrova}, \bibinfo{author}{H.~M. Gibbs},
  \bibinfo{journal}{Nature Materials} \bibinfo{volume}{5}
  (\bibinfo{year}{2006}) \bibinfo{pages}{523–531}.
\bibitem[{Chiodo et~al.(2010)Chiodo, Garc\'{\i}a-Lastra, Iacomino, Ossicini,
  Zhao, Petek, and Rubio}]{Chiodo2010}
\bibinfo{author}{L.~Chiodo}, \bibinfo{author}{J.~M. Garc\'{\i}a-Lastra},
  \bibinfo{author}{A.~Iacomino}, \bibinfo{author}{S.~Ossicini},
  \bibinfo{author}{J.~Zhao}, \bibinfo{author}{H.~Petek},
  \bibinfo{author}{A.~Rubio}, \bibinfo{journal}{Phys. Rev. B}
  \bibinfo{volume}{82} (\bibinfo{year}{2010}) \bibinfo{pages}{045207}.
\bibitem[{Kang and Hybertsen(2010)}]{Kang2010}
\bibinfo{author}{W.~Kang}, \bibinfo{author}{M.~S. Hybertsen},
  \bibinfo{journal}{Phys. Rev. B} \bibinfo{volume}{82} (\bibinfo{year}{2010})
  \bibinfo{pages}{085203}.
\bibitem[{R\"{o}dl and Schleife(2013)}]{Rodl2013}
\bibinfo{author}{C.~R\"{o}dl}, \bibinfo{author}{A.~Schleife},
  \bibinfo{journal}{physica status solidi (a)} \bibinfo{volume}{211}
  (\bibinfo{year}{2013}) \bibinfo{pages}{74–81}.
\bibitem[{Lany(2015)}]{Lany2015}
\bibinfo{author}{S.~Lany}, \bibinfo{journal}{Journal of Physics: Condensed
  Matter} \bibinfo{volume}{27} (\bibinfo{year}{2015}) \bibinfo{pages}{283203}.
\bibitem[{Baldini et~al.(2017)Baldini, Chiodo, Dominguez, Palummo, Moser,
  Yazdi-Rizi, Aub\"{o}ck, Mallett, Berger, Magrez, Bernhard, Grioni, Rubio, and
  Chergui}]{Baldini2017}
\bibinfo{author}{E.~Baldini}, \bibinfo{author}{L.~Chiodo},
  \bibinfo{author}{A.~Dominguez}, \bibinfo{author}{M.~Palummo},
  \bibinfo{author}{S.~Moser}, \bibinfo{author}{M.~Yazdi-Rizi},
  \bibinfo{author}{G.~Aub\"{o}ck}, \bibinfo{author}{B.~Mallett},
  \bibinfo{author}{H.~Berger}, \bibinfo{author}{A.~Magrez},
  \bibinfo{author}{C.~Bernhard}, \bibinfo{author}{M.~Grioni},
  \bibinfo{author}{A.~Rubio}, \bibinfo{author}{M.~Chergui},
  \bibinfo{journal}{Nature Communications} \bibinfo{volume}{8}
  (\bibinfo{year}{2017}).
\bibitem[{Nikl et~al.(2000)Nikl, Bohacek, Mihokova, Kobayashi, Ishii, Usuki,
  Babin, Stolovich, Zazubovich, and Bacci}]{Nikl2000}
\bibinfo{author}{M.~Nikl}, \bibinfo{author}{P.~Bohacek},
  \bibinfo{author}{E.~Mihokova}, \bibinfo{author}{M.~Kobayashi},
  \bibinfo{author}{M.~Ishii}, \bibinfo{author}{Y.~Usuki},
  \bibinfo{author}{V.~Babin}, \bibinfo{author}{A.~Stolovich},
  \bibinfo{author}{S.~Zazubovich}, \bibinfo{author}{M.~Bacci},
  \bibinfo{journal}{Journal of Luminescence} \bibinfo{volume}{87–89}
  (\bibinfo{year}{2000}) \bibinfo{pages}{1136–1139}.
\bibitem[{Laguta et~al.(2007)Laguta, Nikl, and Zazubovich}]{Laguta2007}
\bibinfo{author}{V.~Laguta}, \bibinfo{author}{M.~Nikl},
  \bibinfo{author}{S.~Zazubovich}, \bibinfo{journal}{Radiation Measurements}
  \bibinfo{volume}{42} (\bibinfo{year}{2007}) \bibinfo{pages}{515–520}.
\bibitem[{Pudewill et~al.(1976)Pudewill, Himpsel, Saile, Schwentner, Skibowski,
  Koch, and Jortner}]{Pudewill1976}
\bibinfo{author}{D.~Pudewill}, \bibinfo{author}{F.-J. Himpsel},
  \bibinfo{author}{V.~Saile}, \bibinfo{author}{N.~Schwentner},
  \bibinfo{author}{M.~Skibowski}, \bibinfo{author}{E.~E. Koch},
  \bibinfo{author}{J.~Jortner}, \bibinfo{journal}{The Journal of Chemical
  Physics} \bibinfo{volume}{65} (\bibinfo{year}{1976})
  \bibinfo{pages}{5226–5238}.
\bibitem[{Gr\"{u}nberg and Gabriel(1995)}]{Grunberg1995}
\bibinfo{author}{H.~H.~v. Gr\"{u}nberg}, \bibinfo{author}{H.~Gabriel},
  \bibinfo{journal}{The Journal of Chemical Physics} \bibinfo{volume}{103}
  (\bibinfo{year}{1995}) \bibinfo{pages}{6040–6050}.
\bibitem[{Elliott(1957)}]{Elliott1957}
\bibinfo{author}{R.~J. Elliott}, \bibinfo{journal}{Phys. Rev.}
  \bibinfo{volume}{108} (\bibinfo{year}{1957}) \bibinfo{pages}{1384--1389}.
\bibitem[{Ito and Masumi(1997)}]{Ito1997}
\bibinfo{author}{T.~Ito}, \bibinfo{author}{T.~Masumi},
  \bibinfo{journal}{Journal of the Physical Society of Japan}
  \bibinfo{volume}{66} (\bibinfo{year}{1997}) \bibinfo{pages}{2185–2193}.
\bibitem[{Quintela et~al.(2022)Quintela, Henriques, Tenório, and
  Peres}]{Quintela2022}
\bibinfo{author}{M.~F. C.~M. Quintela}, \bibinfo{author}{J.~C.~G. Henriques},
  \bibinfo{author}{L.~G.~M. Tenório}, \bibinfo{author}{N.~M.~R. Peres},
  \bibinfo{journal}{physica status solidi (b)} \bibinfo{volume}{259}
  (\bibinfo{year}{2022}) \bibinfo{pages}{2200097}.
\bibitem[{Reining(2016)}]{Reining2016}
\bibinfo{author}{L.~Reining}, \bibinfo{title}{{Linear Response and More: the
  Bethe-Salpeter Equation}}, \bibinfo{publisher}{Forschungszentrum Julich (ISBN
  978-3-95806-159-0)}.
\bibitem[{Blase et~al.(2020)Blase, Duchemin, Jacquemin, and Loos}]{Blase2020}
\bibinfo{author}{X.~Blase}, \bibinfo{author}{I.~Duchemin},
  \bibinfo{author}{D.~Jacquemin}, \bibinfo{author}{P.-F. Loos},
  \bibinfo{journal}{The Journal of Physical Chemistry Letters}
  \bibinfo{volume}{11} (\bibinfo{year}{2020}) \bibinfo{pages}{7371–7382}.
\bibitem[{Biswas et~al.(2018)Biswas, Husek, Londo, and Baker}]{Biswas2018}
\bibinfo{author}{S.~Biswas}, \bibinfo{author}{J.~Husek},
  \bibinfo{author}{S.~Londo}, \bibinfo{author}{L.~R. Baker},
  \bibinfo{journal}{Nano Letters} \bibinfo{volume}{18} (\bibinfo{year}{2018})
  \bibinfo{pages}{1228–1233}.
\bibitem[{Varrassi et~al.(2021)Varrassi, Liu, Yavas, Bokdam, Kresse, and
  Franchini}]{Varrassi2021}
\bibinfo{author}{L.~Varrassi}, \bibinfo{author}{P.~Liu}, \bibinfo{author}{Z.~E.
  Yavas}, \bibinfo{author}{M.~Bokdam}, \bibinfo{author}{G.~Kresse},
  \bibinfo{author}{C.~Franchini}, \bibinfo{journal}{Phys. Rev. Mater.}
  \bibinfo{volume}{5} (\bibinfo{year}{2021}) \bibinfo{pages}{074601}.
\bibitem[{Perdew et~al.(1996)Perdew, Burke, and Ernzerhof}]{Perdew1996}
\bibinfo{author}{J.~P. Perdew}, \bibinfo{author}{K.~Burke},
  \bibinfo{author}{M.~Ernzerhof}, \bibinfo{journal}{Physical Review Letters}
  \bibinfo{volume}{77} (\bibinfo{year}{1996}) \bibinfo{pages}{3865--3868}.
\bibitem[{Tran and Blaha(2009)}]{Tran2009}
\bibinfo{author}{F.~Tran}, \bibinfo{author}{P.~Blaha}, \bibinfo{journal}{Phys.
  Rev. Lett.} \bibinfo{volume}{102} (\bibinfo{year}{2009})
  \bibinfo{pages}{226401}.
\bibitem[{Koller et~al.(2012)Koller, Tran, and Blaha}]{Koeller2012}
\bibinfo{author}{D.~Koller}, \bibinfo{author}{F.~Tran},
  \bibinfo{author}{P.~Blaha}, \bibinfo{journal}{Phys. Rev. B}
  \bibinfo{volume}{85} (\bibinfo{year}{2012}) \bibinfo{pages}{155109}.
\bibitem[{Jishi et~al.(2014)Jishi, Ta, and Sharif}]{Jishi2014}
\bibinfo{author}{R.~A. Jishi}, \bibinfo{author}{O.~B. Ta},
  \bibinfo{author}{A.~A. Sharif}, \bibinfo{journal}{J. Phys. Chem. C}
  \bibinfo{volume}{118} (\bibinfo{year}{2014}) \bibinfo{pages}{28344--28349}.
\bibitem[{Ray et~al.(2016)Ray, Himanshu, Lahiri, Kumar, Sen, Bandyopadhyay, and
  Sinha}]{Ray2016}
\bibinfo{author}{R.~Ray}, \bibinfo{author}{A.~K. Himanshu},
  \bibinfo{author}{J.~Lahiri}, \bibinfo{author}{U.~Kumar},
  \bibinfo{author}{P.~Sen}, \bibinfo{author}{S.~K. Bandyopadhyay},
  \bibinfo{author}{T.~P. Sinha}, in: \bibinfo{booktitle}{AIP Conference
  Proceedings}, \bibinfo{publisher}{Author(s)}, \bibinfo{year}{2016}, p.
  \bibinfo{pages}{140041}.
\bibitem[{Mondal et~al.(2018)Mondal, Jha, Himanshu, Lahiri, Singh, Kumar, and
  Ray}]{Mondal2018}
\bibinfo{author}{G.~Mondal}, \bibinfo{author}{D.~Jha}, \bibinfo{author}{A.~K.
  Himanshu}, \bibinfo{author}{J.~Lahiri}, \bibinfo{author}{B.~K. Singh},
  \bibinfo{author}{U.~Kumar}, \bibinfo{author}{R.~Ray}, in:
  \bibinfo{booktitle}{AIP Conference Proceedings},
  \bibinfo{publisher}{Author(s)}, \bibinfo{year}{2018}, p.
  \bibinfo{pages}{090026}.
\bibitem[{Himanshu et~al.(2022)Himanshu, Kumar, Dey, and Ray}]{Himanshu2022}
\bibinfo{author}{A.~Himanshu}, \bibinfo{author}{S.~Kumar},
  \bibinfo{author}{U.~Dey}, \bibinfo{author}{R.~Ray}, \bibinfo{journal}{Physica
  B: Condensed Matter} \bibinfo{volume}{637} (\bibinfo{year}{2022})
  \bibinfo{pages}{413856}.
\bibitem[{Faik et~al.(2012)Faik, Orobengoa, Iturbe-Zabalo, and
  Igartua}]{Faik2012}
\bibinfo{author}{A.~Faik}, \bibinfo{author}{D.~Orobengoa},
  \bibinfo{author}{E.~Iturbe-Zabalo}, \bibinfo{author}{J.~Igartua},
  \bibinfo{journal}{Journal of Solid State Chemistry} \bibinfo{volume}{192}
  (\bibinfo{year}{2012}) \bibinfo{pages}{273--283}.
\bibitem[{esi}]{esi}
\bibinfo{note}{The Supplemental information provides details of the
  crystal structure, XRD pattern, methods, exciton characteristics, further
  details of the electronic and optical properties, and uses additional
  references
  \cite{SahaDasgupta2020,Travis2016,Baerends2013,Draxl2006,Davis1970,Barton1999}}.
\bibitem[{Goldschmidt(1926)}]{Goldschmidt1926}
\bibinfo{author}{V.~M. Goldschmidt}, \bibinfo{journal}{Die Naturwissenschaften}
  \bibinfo{volume}{14} (\bibinfo{year}{1926}) \bibinfo{pages}{477–485}.
\bibitem[{Bartel et~al.(2019)Bartel, Sutton, Goldsmith, Ouyang, Musgrave,
  Ghiringhelli, and Scheffler}]{Bartel2019}
\bibinfo{author}{C.~J. Bartel}, \bibinfo{author}{C.~Sutton},
  \bibinfo{author}{B.~R. Goldsmith}, \bibinfo{author}{R.~Ouyang},
  \bibinfo{author}{C.~B. Musgrave}, \bibinfo{author}{L.~M. Ghiringhelli},
  \bibinfo{author}{M.~Scheffler}, \bibinfo{journal}{Science Advances}
  \bibinfo{volume}{5} (\bibinfo{year}{2019}) \bibinfo{pages}{aav0693}.
\bibitem[{Davis and Mott(1970)}]{Davis1970}
\bibinfo{author}{E.~A. Davis}, \bibinfo{author}{N.~F. Mott},
  \bibinfo{journal}{Philosophical Magazine} \bibinfo{volume}{22}
  (\bibinfo{year}{1970}) \bibinfo{pages}{0903--0922}.
\bibitem[{Blaha et~al.(2018)Blaha, Schwarz, Madsen, Kvasnicka, Luitz,
  Laskowski, Tran, and Marks}]{wien2k}
\bibinfo{author}{P.~Blaha}, \bibinfo{author}{K.~Schwarz},
  \bibinfo{author}{G.~K.~H. Madsen}, \bibinfo{author}{D.~Kvasnicka},
  \bibinfo{author}{J.~Luitz}, \bibinfo{author}{R.~Laskowski},
  \bibinfo{author}{F.~Tran}, \bibinfo{author}{L.~D. Marks},
  \bibinfo{title}{{WIEN2k}, an augmented plane wave {+} local orbitals program
  for calculating crystal properties}, \bibinfo{year}{2018}.
  \bibinfo{note}{(Karlheinz Schwarz, Techn. Universität Wien, Austria)}.
\bibitem[{Blaha et~al.(2020)Blaha, Schwarz, Tran, Laskowski, Madsen, and
  Marks}]{Blaha2020}
\bibinfo{author}{P.~Blaha}, \bibinfo{author}{K.~Schwarz},
  \bibinfo{author}{F.~Tran}, \bibinfo{author}{R.~Laskowski},
  \bibinfo{author}{G.~K.~H. Madsen}, \bibinfo{author}{L.~D. Marks},
  \bibinfo{journal}{The Journal of Chemical Physics} \bibinfo{volume}{152}
  (\bibinfo{year}{2020}) \bibinfo{pages}{074101}.
\bibitem[{Morales-Garcia et~al.(2017)Morales-Garcia, Valero, and
  Illas}]{MoralesGarca2017}
\bibinfo{author}{A.~Morales-Garcia}, \bibinfo{author}{R.~Valero},
  \bibinfo{author}{F.~Illas}, \bibinfo{journal}{The Journal of Physical
  Chemistry C} \bibinfo{volume}{121} (\bibinfo{year}{2017})
  \bibinfo{pages}{18862–18866}.
\bibitem[{Perdew(2009)}]{Perdew2009}
\bibinfo{author}{J.~P. Perdew}, \bibinfo{journal}{International Journal of
  Quantum Chemistry} \bibinfo{volume}{28} (\bibinfo{year}{2009})
  \bibinfo{pages}{497–523}.
\bibitem[{Singh(2010)}]{Singh2010}
\bibinfo{author}{D.~J. Singh}, \bibinfo{journal}{Phys. Rev. B}
  \bibinfo{volume}{82} (\bibinfo{year}{2010}) \bibinfo{pages}{205102}.
\bibitem[{Kim et~al.(2010)Kim, Marsman, Kresse, Tran, and Blaha}]{Kim2010}
\bibinfo{author}{Y.-S. Kim}, \bibinfo{author}{M.~Marsman},
  \bibinfo{author}{G.~Kresse}, \bibinfo{author}{F.~Tran},
  \bibinfo{author}{P.~Blaha}, \bibinfo{journal}{Phys. Rev. B}
  \bibinfo{volume}{82} (\bibinfo{year}{2010}) \bibinfo{pages}{205212}.
\bibitem[{Camargo-Mart\'{\i}nez and Baquero(2012)}]{Martinez2012}
\bibinfo{author}{J.~A. Camargo-Mart\'{\i}nez}, \bibinfo{author}{R.~Baquero},
  \bibinfo{journal}{Phys. Rev. B} \bibinfo{volume}{86} (\bibinfo{year}{2012})
  \bibinfo{pages}{195106}.
\bibitem[{Li et~al.(2013)Li, Walther, Kuc, and Heine}]{Li2013}
\bibinfo{author}{W.~Li}, \bibinfo{author}{C.~F.~J. Walther},
  \bibinfo{author}{A.~Kuc}, \bibinfo{author}{T.~Heine},
  \bibinfo{journal}{Journal of Chemical Theory and Computation}
  \bibinfo{volume}{9} (\bibinfo{year}{2013}) \bibinfo{pages}{2950–2958}.
\bibitem[{Borlido et~al.(2019)Borlido, Aull, Huran, Tran, Marques, and
  Botti}]{Borlido2019}
\bibinfo{author}{P.~Borlido}, \bibinfo{author}{T.~Aull}, \bibinfo{author}{A.~W.
  Huran}, \bibinfo{author}{F.~Tran}, \bibinfo{author}{M.~A.~L. Marques},
  \bibinfo{author}{S.~Botti}, \bibinfo{journal}{Journal of Chemical Theory and
  Computation} \bibinfo{volume}{15} (\bibinfo{year}{2019})
  \bibinfo{pages}{5069–5079}.
\bibitem[{Dresselhaus et~al.(2018)Dresselhaus, Dresselhaus, Cronin, and
  Filho}]{Dresselhaus2018}
\bibinfo{author}{M.~Dresselhaus}, \bibinfo{author}{G.~Dresselhaus},
  \bibinfo{author}{S.~Cronin}, \bibinfo{author}{A.~G.~S. Filho},
  \bibinfo{title}{Solid State Properties}, \bibinfo{publisher}{Springer Berlin
  publishing}, \bibinfo{year}{2018}.
\bibitem[{Wang et~al.(2017)Wang, Daiber, Frost, Mann, Garnett, Walsh, and
  Ehrler}]{Wang2017}
\bibinfo{author}{T.~Wang}, \bibinfo{author}{B.~Daiber}, \bibinfo{author}{J.~M.
  Frost}, \bibinfo{author}{S.~A. Mann}, \bibinfo{author}{E.~C. Garnett},
  \bibinfo{author}{A.~Walsh}, \bibinfo{author}{B.~Ehrler},
  \bibinfo{journal}{Energy {\&} Environmental Science} \bibinfo{volume}{10}
  (\bibinfo{year}{2017}) \bibinfo{pages}{509–515}.
\bibitem[{Yu and Cardona(2010)}]{Yu_book2010}
\bibinfo{author}{P.~Y. Yu}, \bibinfo{author}{M.~Cardona},
  \bibinfo{title}{Fundamentals of Semiconductors},
  \bibinfo{publisher}{Springer-Verlag Heidelberg}, \bibinfo{edition}{4th}
  edition, \bibinfo{year}{2010}.
\bibitem[{Jiang et~al.(2022)Jiang, Jiang, Liu, Lu, and Zhong}]{Jiang2022}
\bibinfo{author}{X.~Jiang}, \bibinfo{author}{B.~Jiang},
  \bibinfo{author}{Y.~Liu}, \bibinfo{author}{J.~Lu},
  \bibinfo{author}{C.~Zhong}, \bibinfo{journal}{The Journal of Physical
  Chemistry Letters} \bibinfo{volume}{13} (\bibinfo{year}{2022})
  \bibinfo{pages}{7306–7313}.
\bibitem[{Ali et~al.(2024)Ali, Khan, Alshgari, Mohammad, and Khandy}]{Ali2024}
\bibinfo{author}{M.~A. Ali}, \bibinfo{author}{A.~Khan}, \bibinfo{author}{R.~A.
  Alshgari}, \bibinfo{author}{S.~Mohammad}, \bibinfo{author}{S.~A. Khandy},
  \bibinfo{journal}{Optical and Quantum Electronics} \bibinfo{volume}{56}
  (\bibinfo{year}{2024}) \bibinfo{pages}{931}.
\bibitem[{Baerends et~al.(2013)Baerends, Gritsenko, and van
  Meer}]{Baerends2013}
\bibinfo{author}{E.~J. Baerends}, \bibinfo{author}{O.~V. Gritsenko},
  \bibinfo{author}{R.~van Meer}, \bibinfo{journal}{Physical Chemistry Chemical
  Physics} \bibinfo{volume}{15} (\bibinfo{year}{2013}) \bibinfo{pages}{16408}.
\bibitem[{Ambrosch-Draxl and Sofo(2006)}]{Draxl2006}
\bibinfo{author}{C.~Ambrosch-Draxl}, \bibinfo{author}{J.~O. Sofo},
  \bibinfo{journal}{Computer Physics Communications} \bibinfo{volume}{175}
  (\bibinfo{year}{2006}) \bibinfo{pages}{1--14}.
\bibitem[{Barton et~al.(1999)Barton, Shtein, Wilson, Soled, and
  Iglesia}]{Barton1999}
\bibinfo{author}{D.~G. Barton}, \bibinfo{author}{M.~Shtein},
  \bibinfo{author}{R.~D. Wilson}, \bibinfo{author}{S.~L. Soled},
  \bibinfo{author}{E.~Iglesia}, \bibinfo{journal}{The Journal of Physical
  Chemistry B} \bibinfo{volume}{103} (\bibinfo{year}{1999})
  \bibinfo{pages}{630--640}.
\bibitem[{Saha-Dasgupta(2020)}]{SahaDasgupta2020}
\bibinfo{author}{T.~Saha-Dasgupta}, \bibinfo{journal}{Materials Research
  Express} \bibinfo{volume}{7} (\bibinfo{year}{2020}) \bibinfo{pages}{014003}.
\bibitem[{Travis et~al.(2016)Travis, Glover, Bronstein, Scanlon, and
  Palgrave}]{Travis2016}
\bibinfo{author}{W.~Travis}, \bibinfo{author}{E.~N.~K. Glover},
  \bibinfo{author}{H.~Bronstein}, \bibinfo{author}{D.~O. Scanlon},
  \bibinfo{author}{R.~G. Palgrave}, \bibinfo{journal}{Chemical Science}
  \bibinfo{volume}{7} (\bibinfo{year}{2016}) \bibinfo{pages}{4548–4556}.

\end{thebibliography}


\clearpage
\newpage
\begin{onecolumngrid}
\setcounter{figure}{0}
\renewcommand\thefigure{S\arabic{figure}}
\setcounter{table}{0}
\renewcommand\thetable{S\Roman{table}}
\setcounter{section}{0}
\setcounter{subsection}{0}

\begin{center}
	\textbf{Supplementary Information}
\end{center}

\section{Crystal Structures}

Fig. \ref{fig:xrd_si} shows the resulting XRD patterns and the corresponding crystal structure parameters are listed in Table \ref{tab:str1}. 
The density functional (DF) calculations correspond to these structures.

\begin{figure}[ht!]
    \centering
    \includegraphics[width=0.95\linewidth]{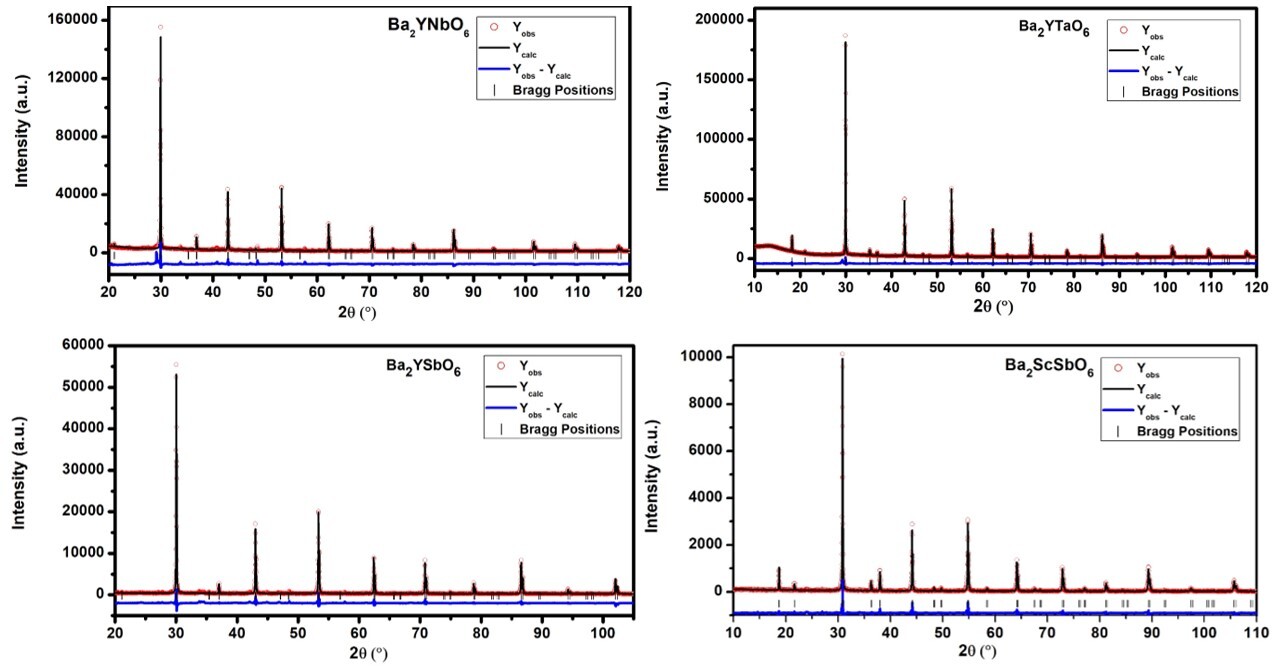}
    \caption{XRD data of the synthesized compoounds.}
    \label{fig:xrd_si}
\end{figure}

\begin{table*}[ht!]
	\caption{Structural details of the synthesized DPOs. The cubic $Fm\bar{3}m$ (No. 225) structures at room temperature are 
	characterized by $a=b=c$ and $\alpha = \beta=\gamma=90^{\circ}$. All the atoms except oxygen occupy high symmetry sites: 
	$A$-site cation at ($1/4$,$1/4$,$1/4$), $B$-site cation at ($0$,$0$,$0$) and $B'$-site cation at ($1/2$,$0$,$0$). Oxygen 
	atoms are centered at ($x$,$0$,$0$). $t$ and $\tau$ are, respectively. The monoclinic structures have $a\ne b \ne c$ and 
	$\beta \neq 90^{\circ}$. $t$ and $\tau$ are, respectively, the Goldschmidt's and Bartel's tolerance factors.}
\label{tab:str1}
    \begin{tabularx}{0.99\textwidth}{ p{2cm} p{1.95cm} p{1.95cm} p{1.95cm} p{1.95cm}  p{1.95cm}  p{1.95cm} p{3.95cm} }
    \hline \hline
        {\bf Param} & {\bf CSS}$^*$ & {\bf SSS}$^\dagger$ & {\bf BSS} & {\bf BYS} & {\bf BYN} & {\bf BYT} & {\bf BST}$^\ddagger$ \\
        \hline \hline
        Space Group \newline (No.)& $P2_{1}n$ & $P2_{1}n$ & $Fm\bar{3}m$  & $Fm\bar{3}m$ & $Fm\bar{3}m$ & $Fm\bar{3}m$  & $Fm\bar{3}m$ \\
        $a$ (\AA) &  5.5088 & 5.6915 & 8.1996 & 8.408(1) & 8.436(8) & 8.438(3) &  8.226 \\
        $b$ (\AA) & 5.6226 & 5.6778 & - & - & - & - & - \\
        $c$ (\AA) & 7.8601 & 8.0244 & - & - & - & - & -\\
        $\beta$ ($^{\circ}$) & 90.02 & 90.03 & 90 & 90 & 90 & 90 & 90\\
        $x$ &  &  & 0.25412 & 0.258(2) & 0.273(2) & 0.260(3) & 0.245 \\
        $Z$ & 2 & 2 & 1 & 1 & 1 & 1 & 1 \\
        Vol. (\AA$^3$/$Z$) & 126.73 & 129.65 & 137.82 & 148.61 & 150.13 & 150.21 & 139.16 \\
        \\
        $B$-O (\AA) & 2.083, \newline 2.093,\newline 2.094 & 2.051, \newline 2.062,\newline 2.072 & 2.084 & 2.171 & 2.305 & 2.198 & 2.097 \\
        $B'$-O (\AA) & 1.992, \newline 1.992,\newline 1.995 & 1.975, \newline 1.992, \newline 2.001 & 2.016 & 2.033 & 1.914 & 2.021 & 2.015 \\
        $\measuredangle B$-O-$B'$ & 148.55$^{\circ}$, 148.85$^{\circ}$, 149.37$^{\circ}$ & 163.53$^{\circ}$, 165.59$^{\circ}$, 166.68$^{\circ}$ & 180$^{\circ}$ & 180$^{\circ}$ & 180$^{\circ}$ & 180$^{\circ}$ & 180$^{\circ}$ \\
        \\
        $t$ & 0.86 & 0.91 & 0.96 & 0.93 & 0.92 & 0.92 & 0.95 \\
        $\tau$ & 4.64 & 3,91 & 3.46 & 3.68 & 3.77 & 3.77 & 3.51 \\
        \hline\hline
        \multicolumn{8}{c}{\textbf{Atomic positions for monoclinic $P2_{1}/n$ (No. 14) structures} } \\
        \hline\hline
        {\bf Compound} & \multicolumn{2}{l}{Sr ($4e$)} & \multicolumn{2}{l}{O$_1$ ($4e$)} & \multicolumn{2}{l}{O$_2$ ($4e$)} & O$_3$ ($4e$) \\
        \hline
        SSS & \multicolumn{2}{l}{(0.0021, 0.0099, 0.2487)} & \multicolumn{2}{l}{(0.2615, 0.2680, 0.0280)} & \multicolumn{2}{l}{(0.2740, 0.2620, 0.4770)} & (0.9493, 0.4944, 0.2456) \\
        CSS & \multicolumn{2}{l}{(0.0150, 0.0461, 0.2490)} & \multicolumn{2}{l}{(0.292, 0.303, 0.052)} & \multicolumn{2}{l}{(0.299, 0.284, 0.444)} & (0.908, 0.468, 0.244) \\
    \hline
    \multicolumn{8}{l}{$^*$ Taken from Ref. \cite{Faik2012}.} \\
    \multicolumn{8}{l}{$^\dagger$ Taken from Ref. \cite{Faik2012}.} \\
    \multicolumn{8}{l}{$^\ddagger$ Taken from Ref. \cite{Himanshu2022}.} \\
    \end{tabularx}
\end{table*}

\section{Tolerance factors}

Traditionally, the Goldschmidt tolerance factor, defined as~\cite{Vasala2015,SahaDasgupta2020,Ray2017,Eng2003,Himanshu2022,Travis2016}: 
\begin{equation}
	t = \frac{r_A + r_X}{\sqrt{2}(\bar{r}_{BB'} + r_X)}\,,
\end{equation}
where $r_A$, $\bar{r}_{BB'}$, and $r_O$ are, respectively, the ionic radii of the A-site cation, average ionic radii of the $B$-site cations, and the ionic radius of the $O$ ion, is a predictor of the perovskite structures and their symmetry. $0.9 \le t \le 1$ suggests a cubic structure while smaller values indicate deviation from the cubic structure. As tabulated in Table \ref{tab:str1}, we find $t>0.92$ for the cubic structures (BSS, BST, BYT, BYS, and BYN) while the monoclinic structures, SSS and CSS have $t=0.91$ and $t=0.86$, respectively, suggesting that the monoclinic distortions in the latter compounds is weak \cite{Ray2017}. 
%

Recently, Bartel {\it et al.} \cite{Bartel2019} proposed a new tolerance factor with higher accuracy in predicting perovskite structures: 
\begin{equation}
    \tau = \frac{r_X}{r_B} - n_A \Big( n_A - \frac{r_A/r_B}{\ln(r_A/r_B)} \Big)\,,
\end{equation}
where $n_A$ is the oxidation state of A, and $r_A > r_B$. For application to double perovskites, average value of the ionic values of the $B$- and $B'$- 
site cations, $\bar{r}_{BB'}$, have been used. $\tau = 4.18$ is considered to be the threshold value such that higher $\tau$ values represent non-perovskite structures. 
With the exception of CSS, we find that all the compounds have $\tau < 4.0$ (see Table \ref{tab:str1}). For CSS, however, we find $\tau = 4.64$.

\clearpage
\clearpage

\section{Optical gaps and excitons modes from the UV-Vis spectrum}

The optical gaps were estimated based on the simple Kubelka-Munk (KM) function without excitons. Fig. \ref{fig:KM_direct_si1} and 
\ref{fig:KM_direct_si2} shows the fit and the corresponding gap values. The estimated gap values are: 4.08 eV for BYN, 4.66 eV for BYT, 
4.67 eV for BYS, 4.57 eV for BST, 4.98 eV for SSS, and 4.74 eV for CSS. It is important to note that the estimated gap for CSS is 
lower than for SSS while density functional calculations suggest that the electronic gaps for CSS should be higher.

\begin{figure}[ht!]
    \includegraphics[scale=1.499,angle=0]{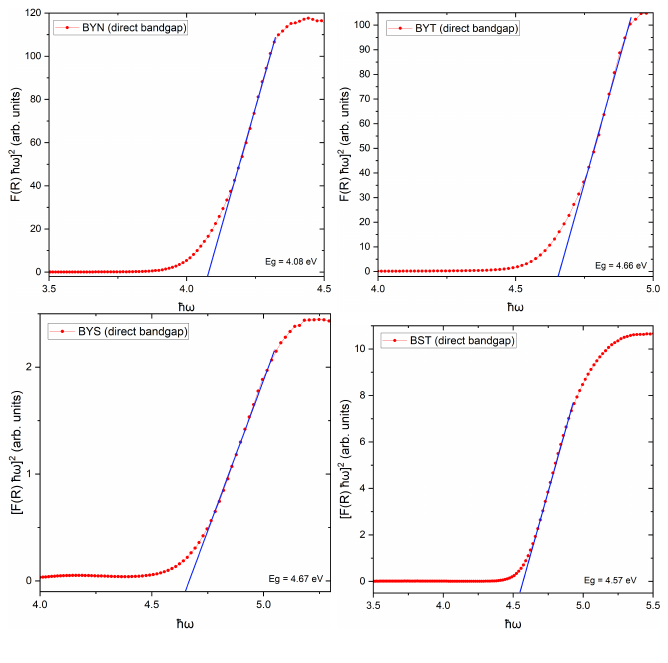}
    \caption{Effective absorption spectra in terms of the Kubelka-Munk function and direct bandgaps of the cubic compounds considered.}
    \label{fig:KM_direct_si1}
\end{figure}

\begin{figure}[ht!]
    \includegraphics[scale=1.499,angle=0]{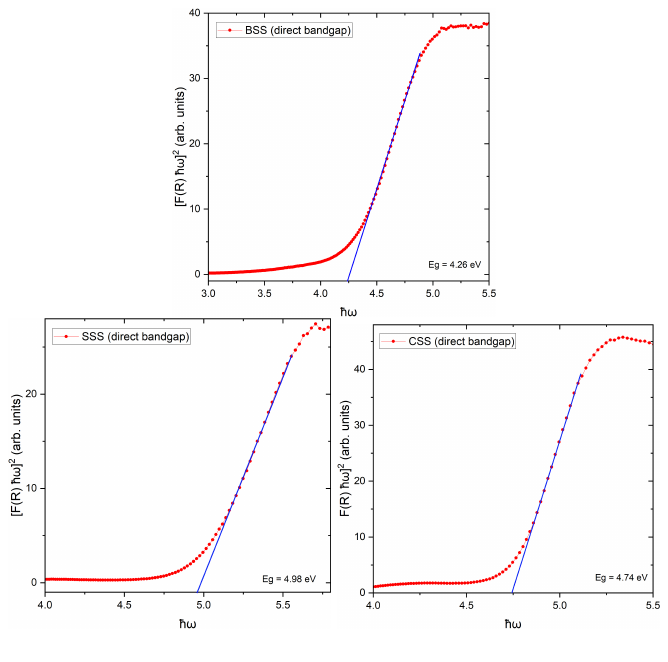}
	\caption{Effective absorption spectra in terms of the Kubelka-Munk function and direct bandgaps of \dpo\ ($A$ = Ba, Sr, Ca).}
    \label{fig:KM_direct_si2}
\end{figure}

Refined estimates for the optical gaps were obtained using a exciton+continuum model (E+C model), as discussed in Methods. 
Fig. \ref{fig:full_fits_si} show the fitting of the low-energy and the high-energy part of the UV-Vis data using the adopted models with excitons. 
The $\chi^2$ values for the E+C model for the high-energy part of the data are: $3.86 \times 10^{-2}$ for BYN, 
$2.27 \times 10^{-2}$ for BYT,  $4.96 \times 10^{-4}$ for BYS,  $1.05 \times 10^{-3}$ for BST, $1.71 \times 10^{-3}$ for SSS, and 
$7.60 \times 10^{-3}$ for CSS. The corresponding gaps are presented in Table I of the main text.

Table \ref{tab:exciton_fit} lists the optimized fit parameters for all the compounds considered in the study. These parameters include 
the exciton binding energy \( E_b \) and the peak linewidth \( \Gamma_n \) for each exciton mode, indicating the variation in exciton 
behavior across the compounds.

\begin{figure}
    \includegraphics[scale=1.5,angle=0]{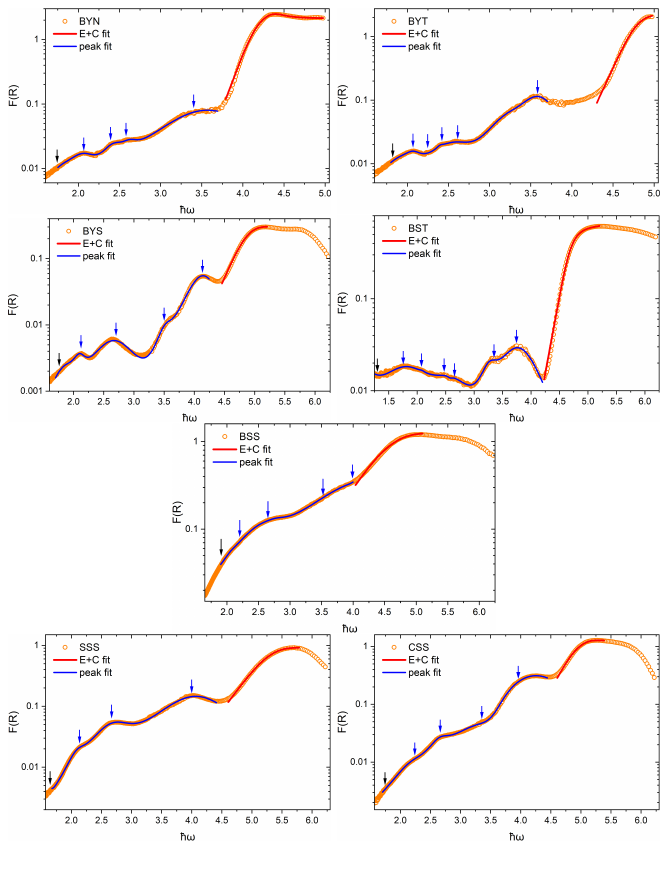}
	\caption{Full absorption spectrum based on the UV-Vis data (symbols) and the correpsonding fits to the low-energy and the high-energy parts of the data (solid lines). The high-energy part of the data is fit using E+C model. In the low-energy parts, arrows indicate the identified exciton peaks. The first peak (black color) is the {\it dummy} peak.}
    \label{fig:full_fits_si}
\end{figure}

\begin{table*}[ht!]
\caption{Details of the fits for each identified exciton peak. Variations in the exciton binding energies \( E_b \) and the peak 
	linewidths \( \Gamma_n \) indicate the variation in exciton behavior across the different double perovskite compounds.}
\label{tab:exciton_fit}
	\def\arraystretch{1.5}
    \begin{tabularx}{0.99\textwidth}{ p{1.00cm} p{0.70cm} p{2.195cm} p{2.195cm} p{2.195cm} p{2.195cm} p{2.195cm} p{2.195cm} p{2.195cm} }  
    \hline \hline
        \multicolumn{2}{l}{\bf Param} & {\bf CSS} & {\bf SSS} & {\bf BSS} & {\bf BYS} & {\bf BYN} & {\bf BYT} & {\bf BST} \\
        \hline \hline
        $n=1$ & $E_b/n^2$ & 2.499 & 2.839 & 2.078 & 2.444 & 2.014 & 2.587 & 2.905 \\
            & $\Gamma$ & 0.196$\pm$0.053 & 0.119$\pm$0.009 & 0.336$\pm$0.014 & 0.079$\pm$0.043 & 0.141$\pm$0.046 & 0.074$\pm$0.041  & 0.388$\pm$0.029 \\
            \\
        $n=2$ & $E_b/n^2$ & 2.077 & 2.311 & 1.628 & 1.970 & 1.688 & 2.427 & 2.585 \\
            & $\Gamma$ & 0.129$\pm$0.012 & 0.272$\pm$0.005 & 0.124$\pm$0.028 & 0.028$\pm$0.006 & 0.106$\pm$0.017 & 0.031$\pm$0.022 & 0.035$\pm$ 0.020 \\
         \\
        $n=3$ & $E_b/n^2$ & 1.387 & 0.980 & 0.756 & 1.032 & 1.500 & 2.226 & 2.187 \\
            & $\Gamma$ & 0.229$\pm$0.006 & 0.475$\pm$0.003 & 0.682$\pm$0.019 & 0.105$\pm$0.004 & 0.086$\pm$0.013 & 0.242$\pm$0.036 & 0.036$\pm$0.016 \\
            \\
        $n=4$ & $E_b/n^2$ & 0.780 & - & 0.29 & 0.524 & 0.677 & 2.058 & 2.005 \\
            & $\Gamma$ & 0.458$\pm$0.004 & - & 0.026$\pm$0.011 & 0.207$\pm$0.001 & 0.440$\pm$0.004 & 0.197$\pm$.012 & 0.321$\pm$0.021 \\
            \\
        $n=5$ & $E_b/n^2$ & - & - & - & - & - & 1.086 & 1.310 \\
            & $\Gamma$ & - & - & - & - & - & 0.192$\pm$0.005 & 0.124$\pm$0.005 \\
            \\
        $n=6$ & $E_b/n^2$ & - & - & - & - & - & - & 0.916 \\
            & $\Gamma$ & - & - & - & - & - & - &  0.287$\pm$0.003 \\
            \\
	    $\chi^2$ & & $5.64 \times 10^{-4}$ & $9.43 \times 10^{-4}$ & $2.76 \times 10^{-4}$ & $4.34 \times 10^{-5}$ & $3.66 \times 10^{-4}$ & $2.84 \times 10^{-4}$ & $8.64 \times 10^{-5}$ \\
	    $\chi^2_{\rm red}$ & & $1.29 \times 10^{-6}$ & $2.10 \times 10^{-6}$ & $8.40 \times 10^{-7}$ & $1.02 \times 10^{-7}$ & $8.40 \times 10^{-7}$ & $1.02 \times 10^{-6}$ & $1.20 \times 10^{-7}$ \\
         \hline \hline
    \end{tabularx}
\end{table*}

\clearpage
\newpage

\section{Electronic \& Optical Properties}

\begin{figure}[ht!]
\includegraphics[width=0.75\textwidth,angle=0]{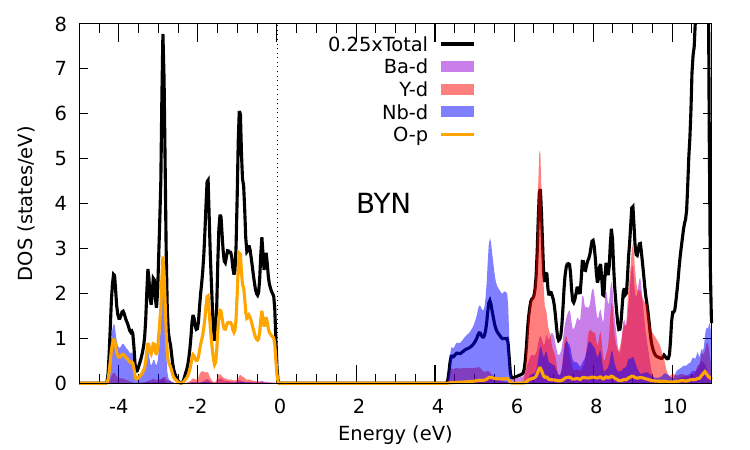}
\caption{Total and atom/orbital-resolved densities of states (DOS) for BYN.}
\label{fig:dos_si_byn}
\end{figure}

\begin{figure}[ht!]
\includegraphics[width=0.95\textwidth,angle=0]{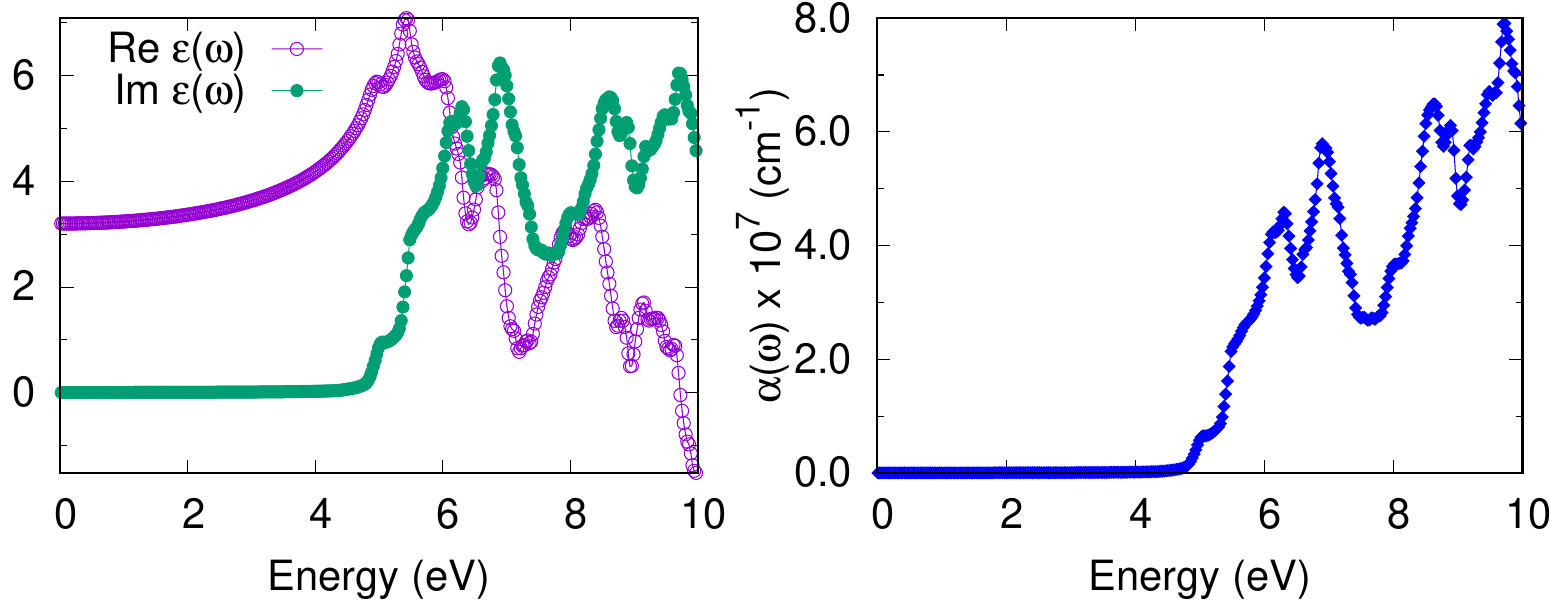}
\caption{(Left) Real and imaginary parts of the dielectric function, (Right) absoprtion coefficient $\alpha(\omega)$ for BYN.}
\label{fig:opti_si_byn}
\end{figure}

\begin{figure}[ht!]
\includegraphics[width=1.0\textwidth,angle=0]{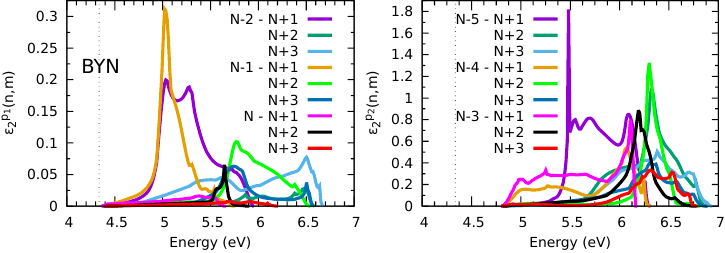}
\caption{Band-resolved $p_1$- and $p_2$-contributions to $\epsilon_2(\omega)$ for BYN, where $N=40$.}
\label{fig:imeps_si_byn}
\end{figure}

\clearpage

\begin{figure}[ht!]
\includegraphics[width=0.75\textwidth,angle=0]{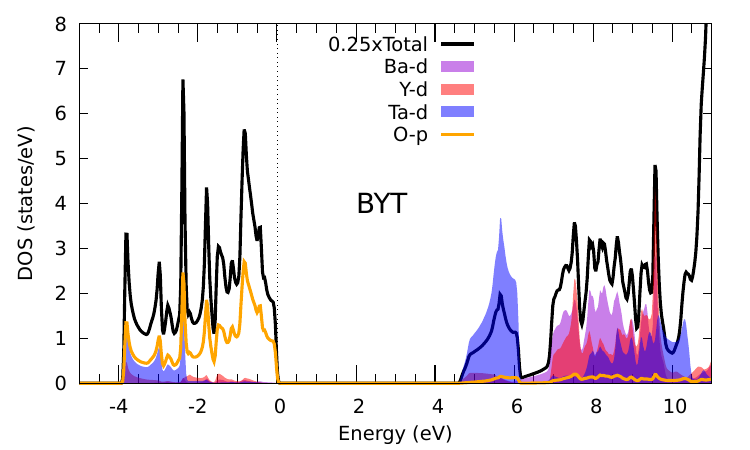}
\caption{Total and atom/orbital-resolved densities of states (DOS) for BYT.}
\label{fig:dos_si_byt}
\end{figure}

\begin{figure}[ht!]
\includegraphics[width=0.95\textwidth,angle=0]{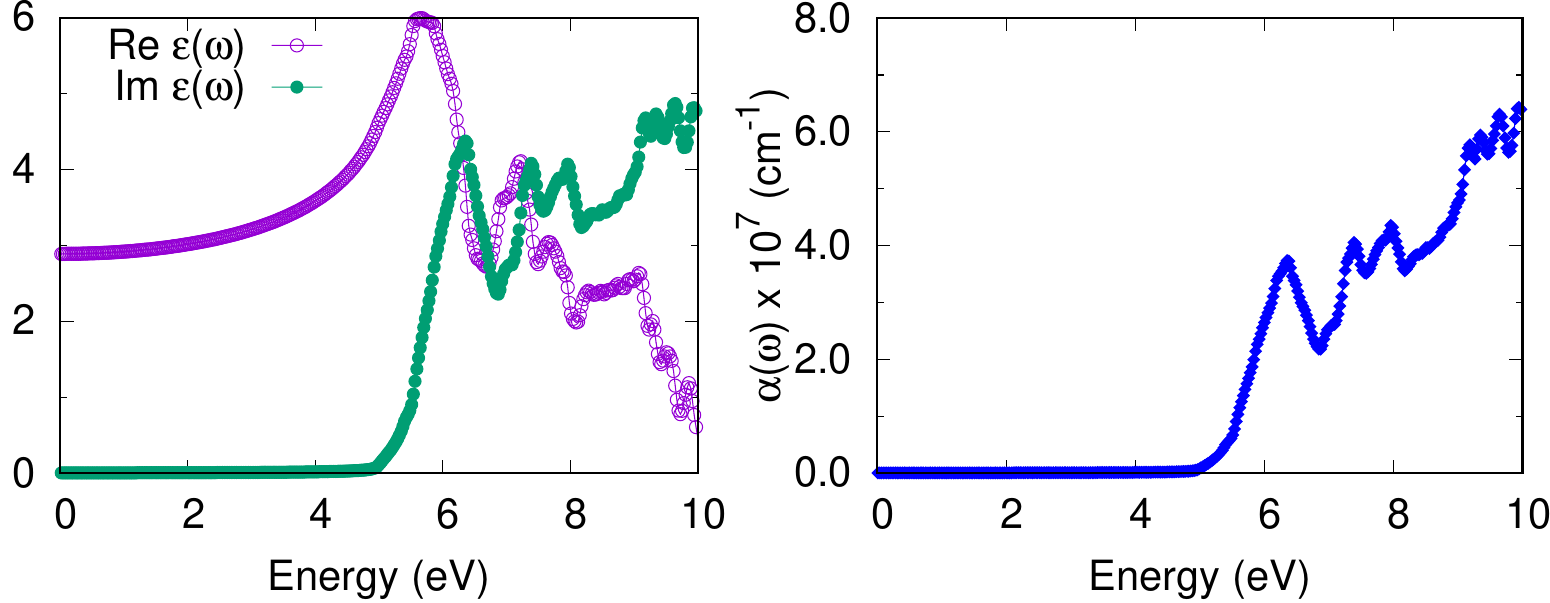}
\caption{(Left) Real and imaginary parts of the dielectric function, (Right) absoprtion coefficient $\alpha(\omega)$ for BYT.}
\label{fig:opti_si_byt}
\end{figure}

\begin{figure}[ht!]
\includegraphics[width=1.0\textwidth,angle=0]{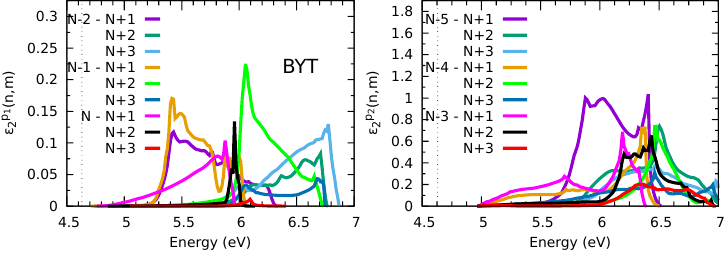}
\caption{Band-resolved $p_1$- and $p_2$-contributions to $\epsilon_2(\omega)$ for BYN.}
\label{fig:imeps_si_byt}
\end{figure}

\clearpage
\clearpage

\begin{figure}[ht!]
\includegraphics[width=0.75\textwidth,angle=0]{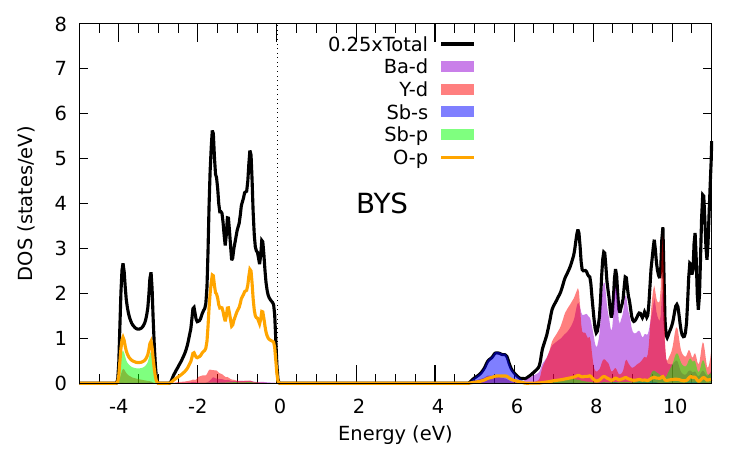}
\caption{Total and atom/orbital-resolved densities of states (DOS) for BYS.}
\label{fig:dos_si_bys}
\end{figure}

\begin{figure}[ht!]
\includegraphics[width=0.95\textwidth,angle=0]{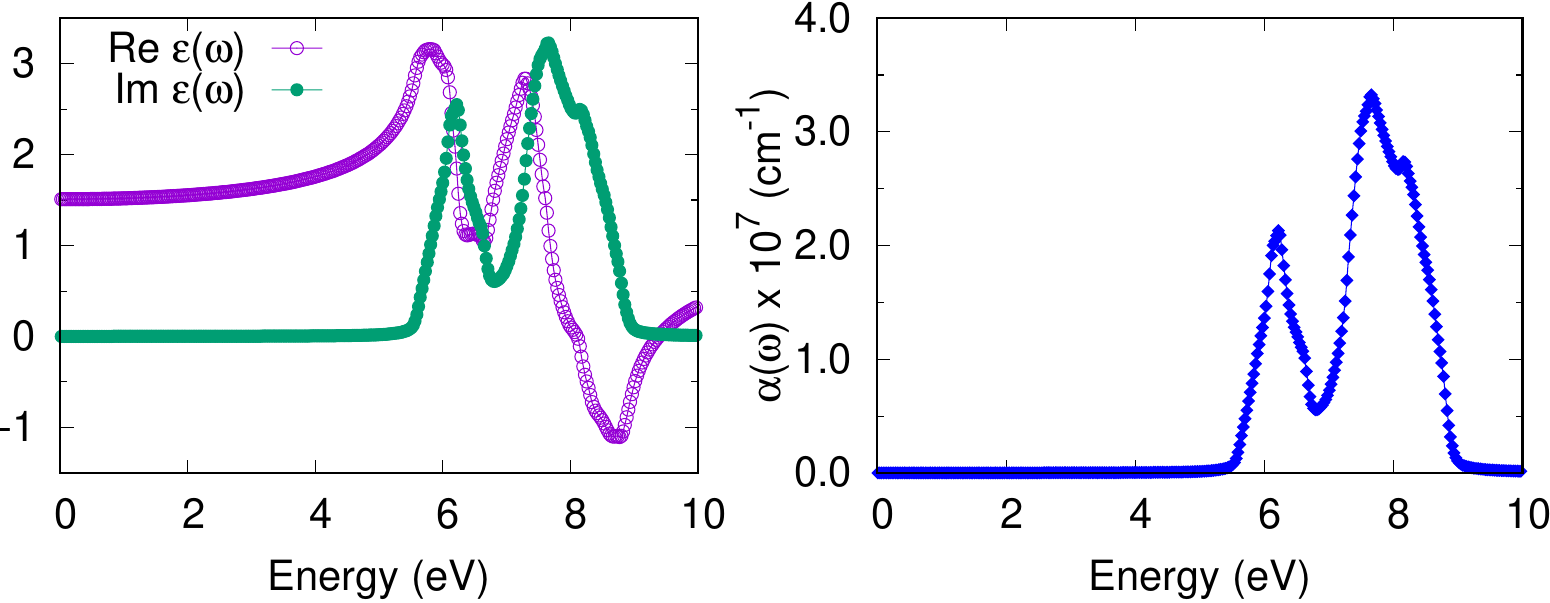}
\caption{(Left) Real and imaginary parts of the dielectric function, (Right) absoprtion coefficient $\alpha(\omega)$ for BYS.}
\label{fig:opti_si_bys}
\end{figure}

\begin{figure}[ht!]
\includegraphics[width=1.0\textwidth,angle=0]{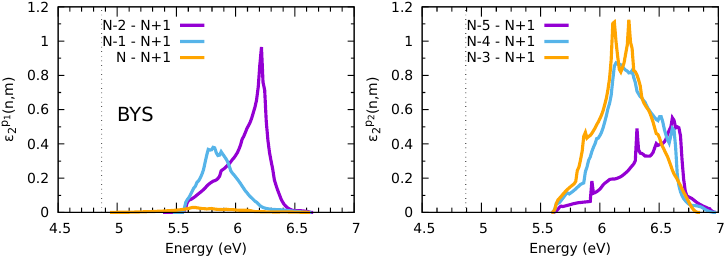}
\caption{Band-resolved $p_1$- and $p_2$-contributions to $\epsilon_2(\omega)$ for BYS.}
\label{fig:imeps_si_bys}
\end{figure}

\clearpage

\begin{figure}[ht!]
\includegraphics[width=0.75\textwidth,angle=0]{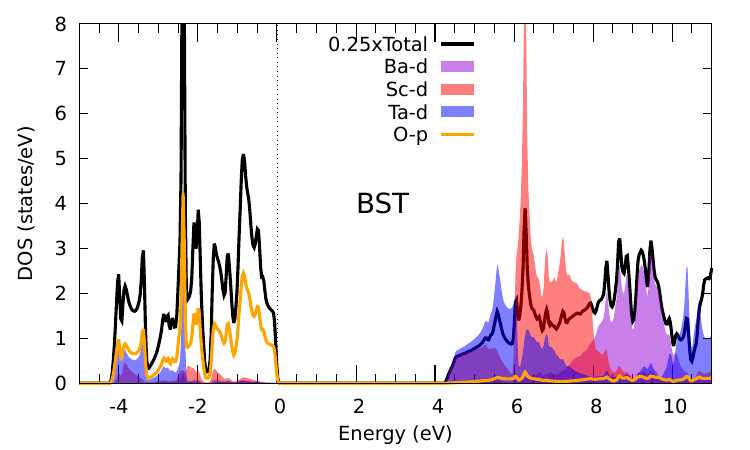}
\caption{Total and atom/orbital-resolved densities of states (DOS) for BST.}
\label{fig:dos_si_bst}
\end{figure}

\begin{figure}[ht!]
\includegraphics[width=0.95\textwidth,angle=0]{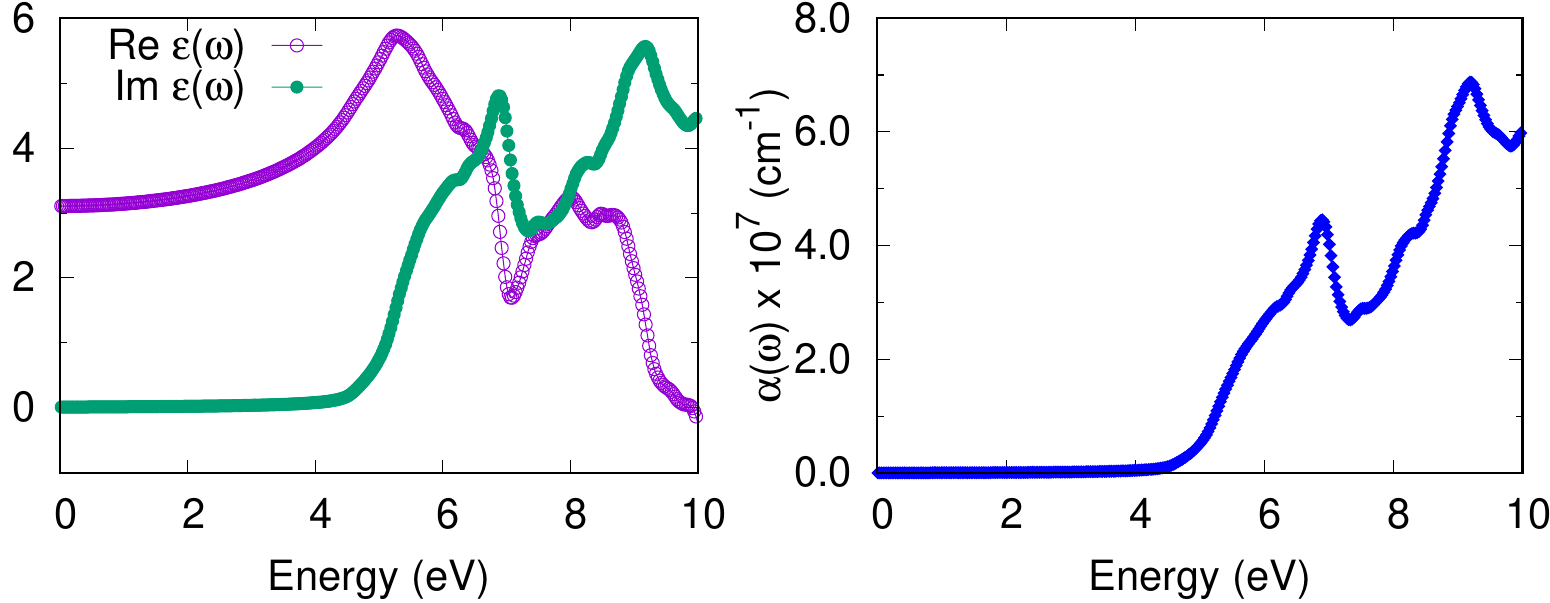}
\caption{(Left) Real and imaginary parts of the dielectric function, (Right) absoprtion coefficient $\alpha(\omega)$ for BST.}
\label{fig:opti_si_bst}
\end{figure}

\begin{figure}[ht!]
\includegraphics[width=1.0\textwidth,angle=0]{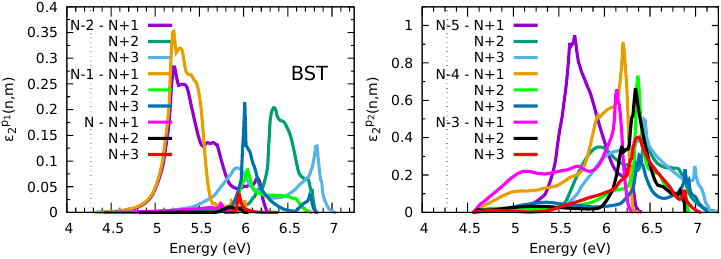}
\caption{Band-resolved $p_1$- and $p_2$-contributions to $\epsilon_2(\omega)$ for BST.}
\label{fig:imeps_si_bst}
\end{figure}

\clearpage

\begin{figure}[ht!]
\includegraphics[width=0.75\textwidth,angle=0]{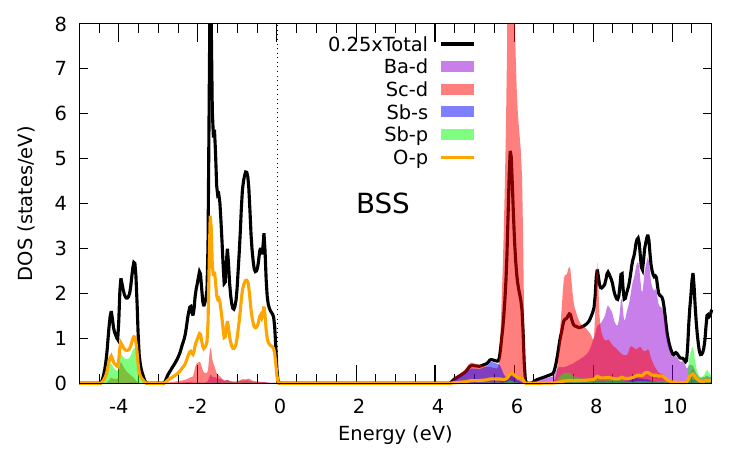}
\caption{Total and atom/orbital-resolved densities of states (DOS) for BSS.}
\label{fig:dos_si_bss}
\end{figure}

\begin{figure}[ht!]
	\includegraphics[width=0.95\textwidth,angle=0]{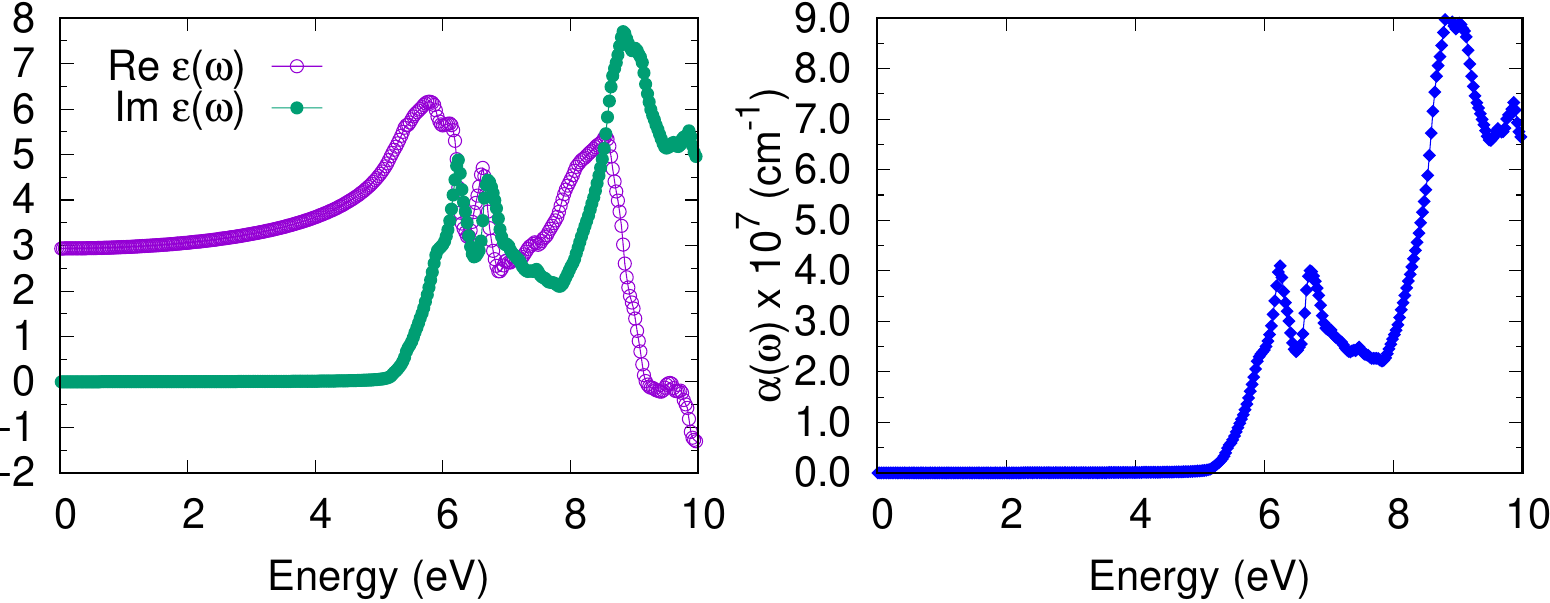}
\caption{(Left) Real and imaginary parts of the dielectric function, (Right) absoprtion coefficient $\alpha(\omega)$ for BSS.}
\label{fig:opti_si_bss}
\end{figure}

\begin{figure}[ht!]
\includegraphics[width=1.0\textwidth,angle=0]{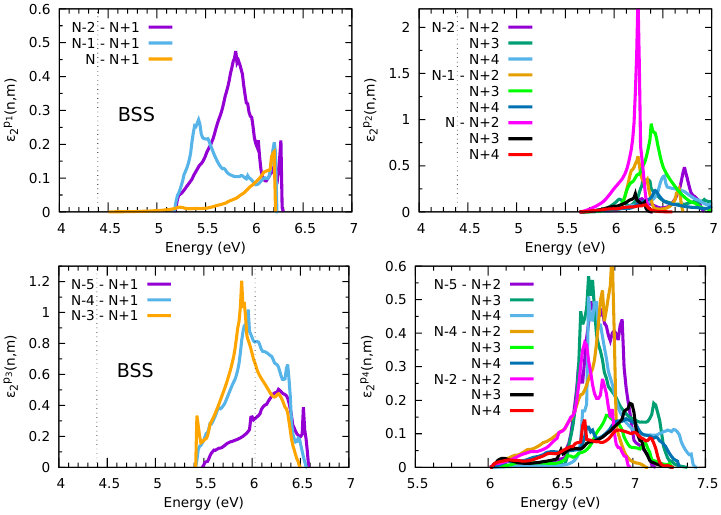}
\caption{Band-resolved $p_1$-, $p_2$-, $p_3$- and $p_4$-contributions to $\epsilon_2(\omega)$ for BSS.}
\label{fig:imeps_si_bss}
\end{figure}

\clearpage

\begin{figure}[ht!]
\includegraphics[width=0.75\textwidth,angle=0]{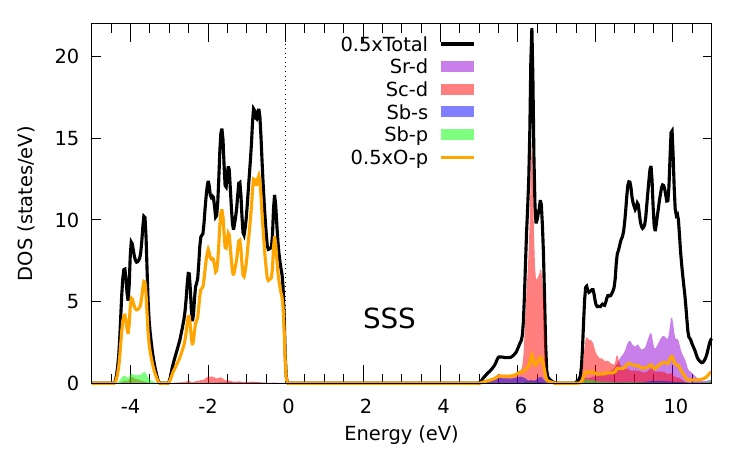}
\caption{Total and atom/orbital-resolved densities of states (DOS) for SSS.}
\label{fig:dos_si_sss}
\end{figure}

\begin{figure}[ht!]
\includegraphics[width=0.95\textwidth,angle=0]{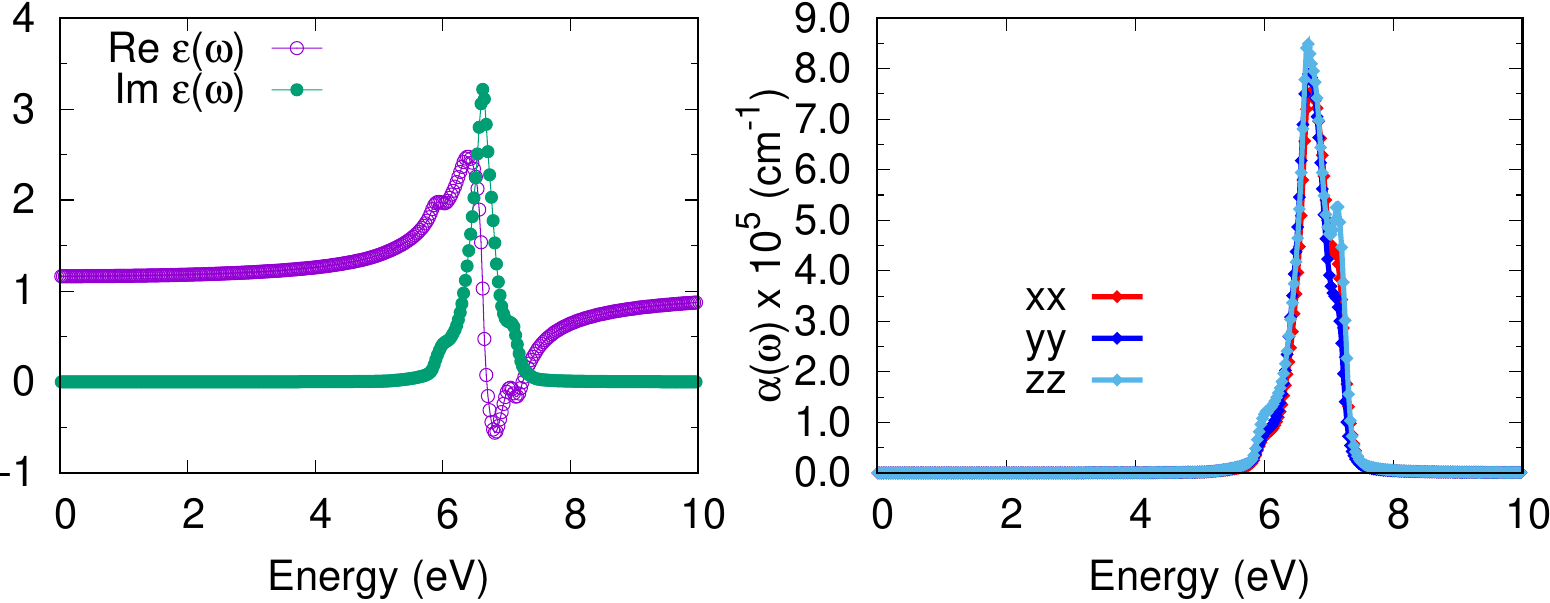}
\caption{(Left) Real and imaginary parts of the dielectric function, (Right) absorption coefficient $\alpha(\omega)$ for SSS. Note the small response values compared to others.}
\label{fig:opti_si_sss}
\end{figure}

\begin{figure}[ht!]
\includegraphics[width=0.95\textwidth,angle=0]{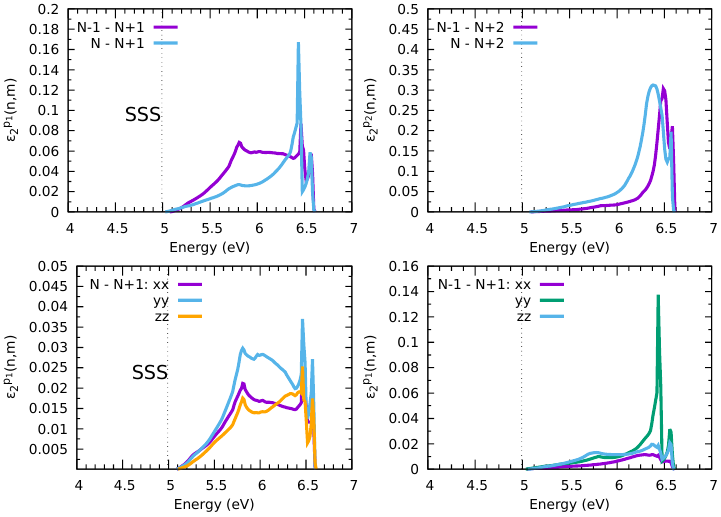}
\caption{(Left) Band-resolved $p_1$ and $p_2$ contributions and direction-resolved contributions for $p_1$ in SSS.}
\label{fig:p1contrib_si_sss}
\end{figure}

\clearpage

\begin{figure}[ht!]
\includegraphics[width=0.75\textwidth,angle=0]{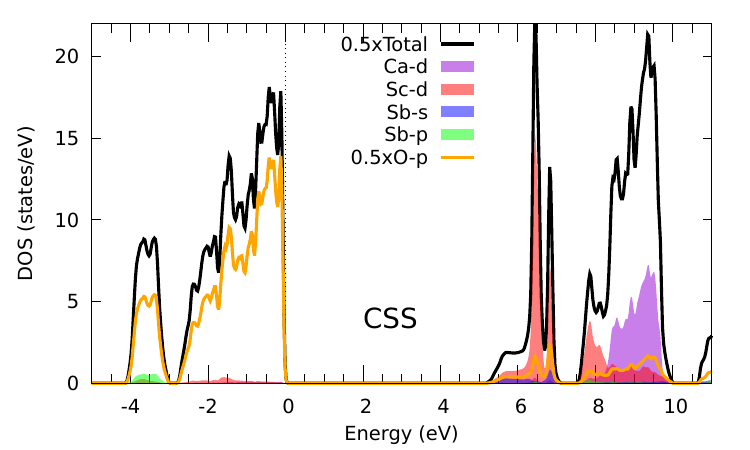}
\caption{Total and atom/orbital-resolved densities of states (DOS) for CSS.}
\label{fig:dos_si_sss}
\end{figure}

\begin{figure}[ht!]
\includegraphics[width=0.95\textwidth,angle=0]{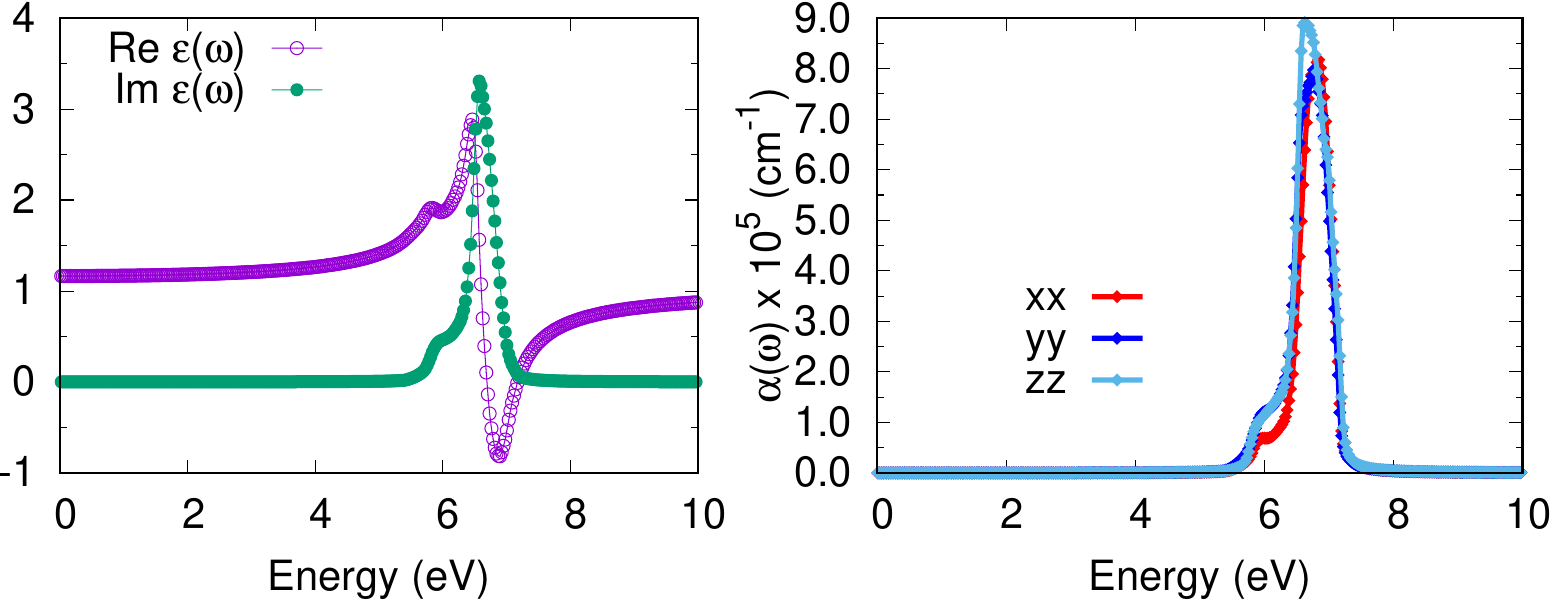}
\caption{(Left) Real and imaginary parts of the dielectric function, (Right) absorption coefficient $\alpha(\omega)$ for CSS. Note the small response values compared to others.}
\label{fig:opti_si_sss}
\end{figure}

\begin{figure}[ht!]
\includegraphics[width=0.95\textwidth,angle=0]{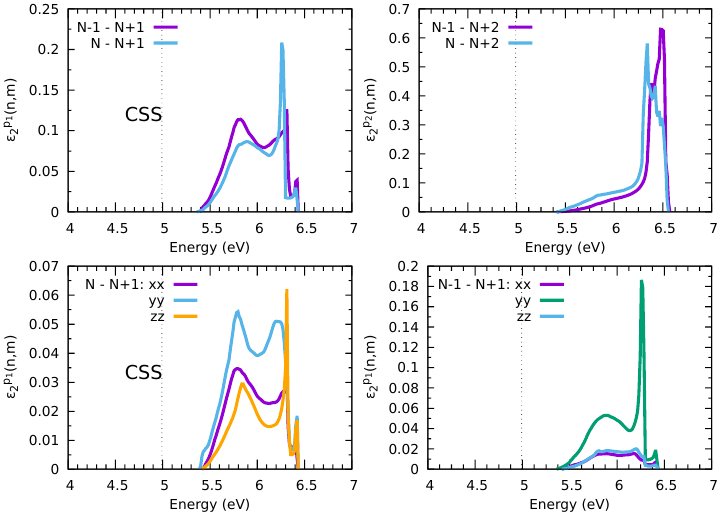}
\caption{(Left) Band-resolved $p_1$ and $p_2$ contributions and direction-resolved contributions for $p_1$ in CSS.}
\label{fig:p1contrib_si_css}
\end{figure}

\clearpage
\newpage
    
\end{onecolumngrid}

\end{document}